\documentclass[sigconf]{acmart}

\AtBeginDocument{%
  }

\setcopyright{none} 
\copyrightyear{2025}
\acmYear{2025}
\acmDOI{XXXXXXX.XXXXXXX}

\acmConference[MICRO 2025]{The 58th IEEE/ACM International Symposium on Microarchitecture}{October 18--22, 2025}{Seoul, Korea}

\acmISBN{978-X-XXXX-XXXX-X/XX/XX}
\renewcommand\footnotetextcopyrightpermission[1]{}

\def\FIGDIR{./figures}
\usepackage{natbib}

\usepackage{amsmath,amssymb,amsfonts,amsthm}
\usepackage{graphicx}
\usepackage{xcolor}
\usepackage{microtype}
\usepackage[italic]{mathastext}
\usepackage{microtype}
\usepackage{mathastext}
\usepackage{libertine}
\usepackage[T1]{fontenc}
\usepackage{textcomp}
\usepackage{zi4}
\usepackage[all]{nowidow}
\usepackage[keeplastbox]{flushend}
\usepackage{fancyhdr}
\usepackage{url}
\usepackage{listings}

\usepackage{makecell}

\usepackage{algorithmic}
\theoremstyle{definition}

\usepackage{newfloat}

\usepackage[small,compact]{titlesec}
\usepackage[linesnumbered]{algorithm2e}
\usepackage{caption}
\usepackage{color}
\usepackage[normalem]{ulem}
\usepackage{xspace}
\usepackage{booktabs} 

\usepackage{makecell}
\usepackage{multirow}
\usepackage{comment}
\usepackage{hhline}
\usepackage[para]{threeparttable} 
\makeatletter
\def\TPT@doparanotes{\par
   \prevdepth\z@ \TPT@hsize
   \TPTnoteSettings
   \parindent\z@ \pretolerance 8
   \linepenalty 200
   \renewcommand\item[1][]{\relax\ifhmode \begingroup
       \unskip
       \advance\hsize 10em 
       \penalty -45 \hskip\z@\@plus\hsize \penalty-19
       \hskip .07\hsize \penalty 9999 \hskip-.15\hsize
       \hskip .01\hsize\@plus-\hsize\@minus.01\hsize 
       \hskip 6em\@plus .3em
      \endgroup\fi
      \tnote{##1}\,\ignorespaces}%
   \let\TPToverlap\relax
   \def\endtablenotes{\par}%
}
\makeatother

\usepackage{hyperref}

\RequirePackage[nameinlink]{cleveref} 
\crefname{chapter}{Chapter}{Chapters}
\crefname{section}{Section}{Sections}
\crefname{subsection}{Subsection}{Subsections}
\crefname{equation}{Equation}{Equations}
\crefname{definition}{Definition}{Definitions}
\crefname{assumption}{Assumption}{Assumptions}
\crefname{theorem}{Theorem}{Theorems}
\crefname{figure}{Figure}{Figures}
\crefname{table}{Table}{Tables}
\crefname{BOX}{Box}{Boxes}
\let\autoref\cref 

\def\FIGDIR{./figures}

\newcommand{\insertFigure}[2]{
    \begin{figure}[t]
        \centering
        \includegraphics[width=\linewidth]{\FIGDIR/#1.pdf}
        \vspace{-7mm}
        \caption{\small #2}
        \vspace{-3mm}
        \label{fig:#1}
    \end{figure}
}

\newcommand{\insertFigureShrink}[3]{
    \begin{figure}[t]
        \centering
        \includegraphics[width=#3\linewidth]{\FIGDIR/#1.pdf}
        \vspace{-4mm}
        \caption{\small #2}
        \vspace{-5mm}
        \label{fig:#1}
    \end{figure}
}

\newcommand{\insertFigureShrinkCustomMargin}[5]{
    \begin{figure}[t]
        \centering
        \includegraphics[width=#3\linewidth]{\FIGDIR/#1.pdf}
        \vspace{#4 mm}
        \caption{\small #2}
        \vspace{#5 mm}
        \label{fig:#1}
    \end{figure}
}

\newcommand{\insertFigureScaleTight}[3]{
    \begin{figure}[t]
        \centering
        \includegraphics[width=#3\linewidth]{\FIGDIR/#1.pdf}
        \vspace{-4mm}
        \caption{\small #2}
        \vspace{-5mm}
        \label{fig:#1}
    \end{figure}
}

\newcommand{\insertWideFigureShrink}[3]{
    \begin{figure*}[h]
        \centering
        \includegraphics[width=#3\textwidth]{\FIGDIR/#1.pdf}
        \vspace{-4mm}
        \caption{\small #2}
        \vspace{-4mm}
        \label{fig:#1}
    \end{figure*}
}

\newcommand{\insertWideFigureShrinkCustomMargin}[5]{
    \begin{figure*}[h]
        \centering
        \includegraphics[width=#3\textwidth]{\FIGDIR/#1.pdf}
        \vspace{#4 mm}
        \caption{\small #2}
        \vspace{#5 mm}
        \label{fig:#1}
    \end{figure*}
}

\DeclareFloatingEnvironment[
  fileext=lob,
  listname={List of Boxes},
  name=Box,
  placement=htp,
]{BOX}

\ifx\forsubmission\undefined
\newcommand{\TODO}[1]{\textcolor{red}{TODO: #1}}
\newcommand{\FT}[1]{\textcolor{orange}{FT: #1}}
\newcommand{\HK}[1]{\textcolor{blue}{HK: #1}}

\else
\newcommand{\TODO}[1]{\textcolor{red}{}}
\newcommand{\CM}[1]{\textcolor{magenta}{}}
\newcommand{\FT}[1]{\textcolor{orange}{}}
\newcommand{\HK}[1]{\textcolor{blue}{}}
\newcommand{\MP}[1]{\textcolor{teal}{}}

\fi

\newcommand{\squishlist}{
 \begin{list}{$\bullet$}
  { \setlength{\itemsep}{0pt}
     \setlength{\parsep}{3pt}
     \setlength{\topsep}{3pt}
     \setlength{\partopsep}{0pt}
     \setlength{\leftmargin}{1.5em}
     \setlength{\labelwidth}{1em}
     \setlength{\labelsep}{0.5em} } }

\newcommand{\squishlisttwo}{
 \begin{list}{$\bullet$}
  { \setlength{\itemsep}{0pt}
     \setlength{\parsep}{0pt}
    \setlength{\topsep}{0pt}
    \setlength{\partopsep}{0pt}
    \setlength{\leftmargin}{2em}
    \setlength{\labelwidth}{1.5em}
    \setlength{\labelsep}{0.5em} } }

\newcommand{\squishend}{
  \end{list}  }

\newcommand{\betterparagraph}[1]{\noindent \textbf{#1. }}

\crefname{algorithm}{Algorithm}{Algorithms}
\crefname{listing}{Code}{Code}

\definecolor{codegreen}{rgb}{0,0.6,0}
\definecolor{codegray}{rgb}{0.5,0.5,0.5}
\definecolor{codepurple}{rgb}{0.58,0,0.82}
\definecolor{backcolour}{rgb}{0.95,0.95,0.92}

\lstdefinestyle{mystyle}{
    backgroundcolor=\color{backcolour},   
    commentstyle=\color{codegreen},
    keywordstyle=\color{magenta},
    numberstyle=\tiny\color{codegray},
    stringstyle=\color{codepurple},
    breakatwhitespace=false,         
    breaklines=true,                 
    captionpos=b,                    
    keepspaces=true,                 
    numbers=left,                    
    numbersep=5pt,                  
    showspaces=false,                
    showstringspaces=false,
    showtabs=false,                  
    tabsize=2,
    basicstyle=\tiny
}

\newcommand{\nickname}{\textsc{FlexiBit}\xspace}

\newcommand{\accelerator}{\textsc{FlexiBit}\xspace}

\newcommand{\ourtree}{\textsc{FBRT}\xspace}

\begin{document}

\title{FlexiBit: Fully Flexible Precision Bit-parallel Accelerator Architecture for Arbitrary Mixed Precision AI}


\author{Faraz Tahmasebi}
\affiliation{%
  \institution{University of California, Irvine}
  \department{Electrical Engineering and Computer Science}
  \city{Irvine}
  \state{CA}
  \country{USA}
}
\email{tahmasef@uci.edu}

\author{Yian Wang}
\authornote{Both authors contributed equally to this research.}
\affiliation{%
  \institution{University of California, Irvine}
  \department{Electrical Engineering and Computer Science}
  \city{Irvine}
  \state{CA}
  \country{USA}
}
\email{yianw11@uci.edu}

\author{Benji Y.H. Huang}
\authornotemark[1]
\affiliation{%
  \institution{University of California, Irvine}
  \department{Electrical Engineering and Computer Science}
  \city{Irvine}
  \state{CA}
  \country{USA}
}
\email{yuhuah2@uci.edu}

\author{Hyoukjun Kwon}
\affiliation{%
  \institution{University of California, Irvine}
  \department{Electrical Engineering and Computer Science}
  \city{Irvine}
  \state{CA}
  \country{USA}
}
\email{hyoukjun.kwon@uci.edu}


\keywords{ML Accelerator, Large Language Models, Quantization, Flexible Precision, Mixed Precision}

\begin{abstract}
Recent research has shown that large language models (LLMs) can utilize low-precision floating point (FP) quantization to deliver high efficiency while maintaining original model accuracy. In particular, recent works have shown the effectiveness of non-power-of-two precisions, such as FP6 and FP5, and diverse sensitivity to low-precision arithmetic of LLM layers, which motivates mixed precision arithmetic including non-power-of-two precisions in LLMs. Although low-precision algorithmically leads to low computational overheads, such benefits cannot be fully exploited due to hardware constraints that support a limited set of power-of-two precisions (e.g., FP8, 16, 32, and 64 in NVIDIA H100 Tensor Core). In addition, the hardware compute units are designed to support standard formats (e.g., E4M3 and E5M2 for FP8). Such practices require re-designing the hardware whenever new precision and format emerge, which leads to high hardware replacement costs to exploit the benefits of new precisions and formats. 


Therefore, in this paper, we propose a new accelerator architecture, \nickname, which efficiently supports FP and INT arithmetic in arbitrary precisions and formats. Unlike previous bit-serial designs, which also provide flexibility but at the cost of performance due to its bit-wise temporal processing nature, \nickname's architecture enables bit-parallel processing of any precision and format without compute unit underutilization. \nickname's new capability to exploit non-power of two precision and format led to 1.66$\times$ and 1.62$\times$ higher performance per area on GPT-3 in FP6 targeting a cloud-scale accelerator, compared to a Tensor Core-like architecture and a state-of-the-art bit-parallel flexible precision accelerator, BitFusion, respectively. Also, the bit-parallel nature of \nickname's architecture led to 3.9 $\times$ higher performance per area compared to a state-of-the-art bit-parallel architecture, which demonstrates the significant benefits of \nickname for emerging precision and formats.

\end{abstract}

\maketitle
\pagestyle{plain}

\section{Introduction}
\label{sec:intro}

Artificial intelligence (AI) models today drive many innovative applications such as autonomous driving~\cite{Cui_2024_WACV}, augmented and virtual reality (AR/VR)~\cite{kwon2023xrbench} and AI chatbot~\cite{openai2023gpt4}, delivering high-quality outputs.
However, recent AI models, such as large language models (LLMs), involve extremely heavy computation and energy due to their vast model sizes.
For example, the smallest Llama2-7B involves $3.86 \times 10^{12}$ FLOPS and 157 Billion parameters, and GPT-3 involves $1.33 \times 10^{14}$ FLOPS and 174 Billion parameters.
This motivated many algorithmic (i.e., model compression~\cite{MLSYS2024_42a452cb}) and hardware innovations (i.e., accelerators~\cite{hpca2021herald}) to enhance the computational performance and efficiency.
Among such optimizations, quantization is one of the most widely adopted approaches, which reduces the bit precision of individual data for lightweight computation and data movement.
Unlike convolutional neural networks (CNNs) mainly quantize data into fixed point data (i.e., integer format), floating point (FP) quantization and mixed precision quantization with FP/INT data (e.g., GPTQ~\cite{frantar2022gptq}) are more dominant in the large LLM domain for high output quality (e.g., accuracy and perplexity)~\cite{micikevicius2022fp8, xia2024fp6}.
Accordingly, the latest hardware such as NVIDIA H100~\cite{nvidia_h100} and GB200~\cite{nvidia_gb200} added hardware support for FP8 format.

Although the hardware is evolving to support low-precision floating point data types, the hardware support is still limited to the power-of-two precision and specific formats under each power-of-two precision (e.g., E5M2\footnote{EXMY indicates that X and Y bits are used for exponent and mantissa, respectively} and E4M3 formats in FP8).
Such hardware-originated constraints limit the benefit from emerging quantization techniques such as the non-power of two FP quantization (e.g., FP6 and FP5~\cite{xia2024fp6}).
Also, some state-of-the-art LLM quantization works explore variants of FP data precision (e.g., E2M1, E1M2, and E3M0 in FP4-LLM~\cite{liu_llm_fp4_2023}), which demands more flexibility to the hardware for exploiting the benefits of such quantization.
~\autoref{fig:Motivation} illustrates such challenges in existing hardware.

\insertWideFigureShrink{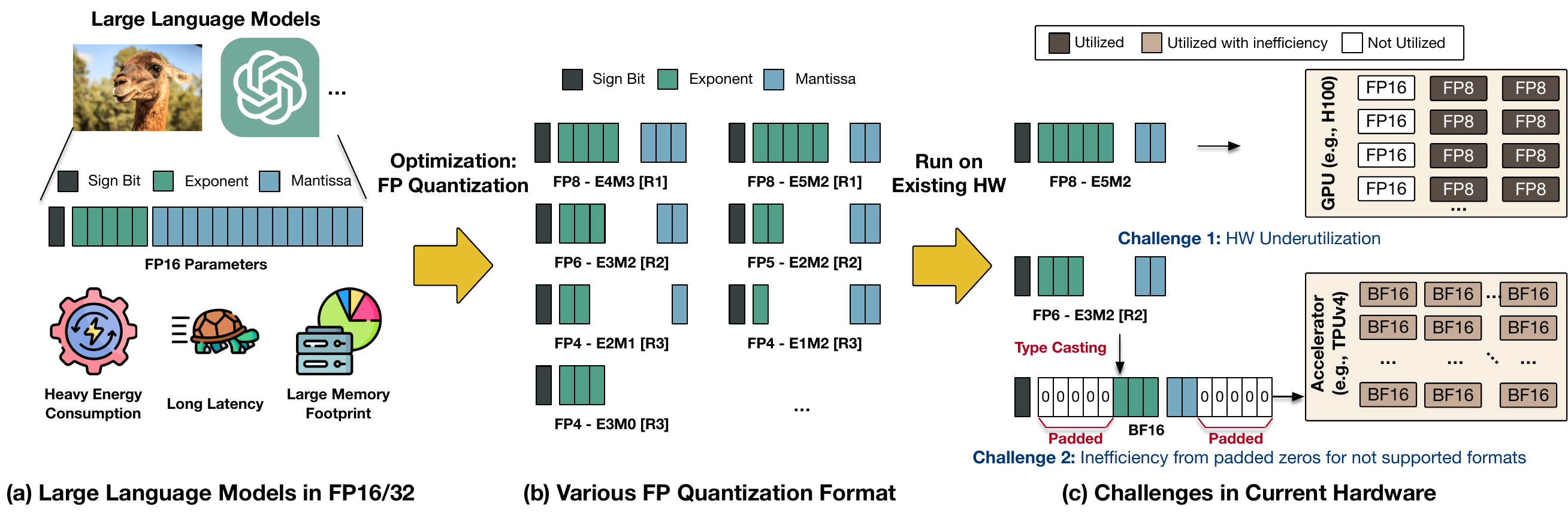}{Various FP quantization formats for large language models (R1: ~\cite{micikevicius2022fp8} R2: \cite{xia2024fp6}, R3: ~\cite{liu_llm_fp4_2023}) and challenges in current GPUs and accelerators for supporting such diverse FP formats. The padding example in (c) is for a positive number.}{1}

The FP format flexibility has a significant impact on the accelerator's performance and energy.
For example, running a model with FP6 (E3M2) quantization on a fixed FP16 processing unit (E5M10) reduces logic utilization by 73\% when FP6 data are padded to fit into FP16, as illustrated in~\autoref{fig:Motivation} (c).
Such a challenge is common for fixed point (INT) data types. To eliminate this underutilization, some works approach the problem with bit-serial architectures\cite{cambricon-p},\cite{Jo_bit-serial}, \cite{MLSYS2022_bit-serial}, \cite{judd2016stripes}, \cite{electronics_bit-serial}. 
While bit-serial processing can adapt to different precisions, their performance is often limited due to the nature of bit-wise temporal processing, which makes it less ideal for recent large models such as LLMs.
Some previous works have proposed bit-parallel architectures that support power-of-two precisions for integer arithmetic \cite{sharma2018bit}, \cite{mix_gemm}, \cite{vlsi_dat_flex_prec}, \cite{2_in_1_Accelerator}, \cite{rapid} \cite{noh2023flexblock}, \cite{aim}, \cite{nvidia_tensorcore}.
However, none of them support (1) FP operations and (2) provide full flexibility of precision for both input and weight.

Therefore, in this work, we explore a new hardware architecture that can support \textit{arbitrary} bit precision and FP data format variants to eliminate (1) the hardware constraints blocking algorithmic innovations in quantization and (2) the need for replacing hardware for new data precision and formats.
One key architecture of our accelerator, \accelerator, is Flexible Bit Reduction Tree (\ourtree) illustrated in~\autoref{fig:Walkthrough} (d), which extends the MAERI~\cite{kwon2018maeri}'s augmented reduction tree toward the bit level. \ourtree includes a configurable shifting mechanism and additional links in the tree architecture for flexible precision/format bit-level arithmetic in a bit-parallel manner. Using the flexibility, \accelerator's multiplier implemented using \ourtree eliminates hardware logic underutilization from data and hardware precision mismatch. We also propose a new flexible-bit adder for exponent addition and data layout management modules, which are crucial for efficient arbitrary precision data processing.


Our evaluation shows that FP format flexibility unlocks significant benefits from non-power-of-two quantization and flexible formats.
Compared to a Tensor Core~\cite{nvidia_tensorcore}-like architecture, \accelerator provides 59\% less latency and 66\% less energy, on average when running FP6 arithmetic, across recent language models we evaluate including Bert-base \cite{kenton2019bert}, Llama2-7B~\cite{touvron2023llama}, Llama-2-70b~\cite{touvron2023llama}, and GPT-3\cite{GPT3}.
Compared to the state-of-the-art flexible precision architecture, Bit-fusion~\cite{sharma2018bit} (extended for FP support) with our extension for floating point arithmetic, \accelerator provided 31\% less latency and 33\% energy, on average, for the same workload.
While a design configuration of \accelerator requires only 0.5\% and 1\%  more hardware and area compared to Tensor Core and Bit-fusion(adopted for FP arithmetic operations), respectively, our analysis of the performance per area shows that the benefits from the extra area surpass the cost, with 20\% and 10\% higher performance per area, on average for cloud-scale accelerator configurations, which demonstrates that the benefits of \accelerator are significantly larger than the costs.

We summarize our contribution as follows:

\squishlist
    {\item We highlight the challenges from recent non-trivial FP/INT quantization methods to the hardware architecture.}
    {\item We develop a fully flexible FP/INT precision/format accelerator architecture, \accelerator, which can efficiently support arbitrary FP/INT precision/format and completely eliminate hardware-oriented constraints on FP/INT quantization algorithms.}
    {\item We evaluate our architecture on recent language models and analyze the benefits over state-of-the-art flexible-precision accelerator architecture.}
\squishend

\section{Background and Motivation}
\label{sec:background}

To discuss technical details and motivation of \accelerator, we first provide an introduction to floating point arithmetic, discuss the recent trend of adopting non-trivial floating point number precision in language models, and provide an insight into current hardware support for arbitrary floating point precision support.
%


\subsection{Floating Point Representation and Arithmetic}
\label{subsec:fp_arithmetic}


%
%
%
%

%
For instance, in FP8 precision, we can distribute 4 bits to the exponent and 3 bits to the mantissa (less sign bit) less 1 bit for sign, which constructs the e4m3 format. 
Alternatively, we can distribute 5 and 2 bits for exponent and mantissa, respectively, which constructs the e5m2 format.
~\autoref{fig:Motivation} (b) illustrates such variants from FP quantization works~\cite{micikevicius2022fp8, xia2024fp6, liu_llm_fp4_2023}.
Here, we discuss the multiplication and addition operations with standard FP encoding as the dominant operations in ML workloads~\cite{chen2016eyeriss_isca}.
%


\betterparagraph{Multiplication}
Multiplication consists of (1) adding exponents, (2) determining sign, (3) multiplying mantissas, (4) shift/truncate mantissa to normalize the results for output precision, and (5) adjusting exponent based on the step 4 results, as the following equation represents:
\begin{equation*}
    A \times B = (-1)^{s_A \oplus s_B} \cdot (1.0 + 0.m_A) \cdot (1.0 + 0.m_B) \cdot 2^{e_A + e_B - \text{bias}}
\end{equation*}
where $S_A$ and $S_B$ are sign bits of A and B, $m_A$ and $m_B$ are mantissa bits, and $e_A$ and $e_B$ are exponent bits.

\betterparagraph{Addition} 
The following equations describe the process of adding two floating point numbers, A and B, by shifting A and taking B as the reference scale.
%

\vspace{-4mm}
\begin{equation*}
\begin{aligned}
    \Delta &= e_B - e_A \\
    m_A' &= 0.m_A << \Delta \\ 
    A' &= (-1)^{s_A} \times (1.0 + 0.m'_A) \times 2^{e_A - \text{bias} + \Delta} &
    \\
    B &= (-1)^{s_B} \times (1.0 + 0.m_B) \times 2^{e_B - \text{bias}}
    \\
    S_O &= (s_A \oplus s_B)? ((max(m'_A, m_B) == m'_A)? s_A : s_B) : s_A   \\ 
    A + B &= A' + B = (-1)^{s_O} \times ((-1)^{s_A}\times 1.m'_A + (-1)^{s_B}) \times 1.m_B  \times 2^{e_B - \text{bias}}
    \\
\end{aligned}
\end{equation*}

\betterparagraph{Micro-Scaling (MX) format Arithmetic}
Micro-Scaling (MX) data format \cite{MX_format} is a precision-adaptive representation designed to optimize storage and computational efficiency in machine learning and high-performance computing workloads. It leverages variable bit-widths for encoding values, where the precision adapts based on the range and accuracy requirements of the data. 
MAC operation of two vectors of $A$ and $W$ in MX-Format follows the following equation:

\vspace{-4mm}
\begin{equation*}
C = \text{Dot}(A, W) = X(A) \cdot X(B) \sum_{i=1}^{i=K} \left(P_i(A) \times P_i(W)\right)
\end{equation*}
\vspace{-3mm}

where $X(A)$ and $X(B)$ are the scaling factor for one block of elements with size K, and $P_i(A)$ and $P_i(A)$ are private (unique) elements. Data precision of $X$, $P$, and $K$ can vary depending on the desired precision.

\subsection{Floating Point Quantization for Language Models}
\label{subsec:precision_llm}
%

%

Due to the difficulty of integer quantization of language models, the mainstream for the language model compression is targeting floating point quantization~\cite{micikevicius2022fp8, xia2024fp6} or mixed precision hybrid approach (e.g., INT4 weight and FP16 activation)~\cite{frantar2022gptq} for high model performance (e.g., perplexity and accuracy).
Among such works, FP6-LLM~\cite{xia2024fp6} reported that FP5 and FP6 provide good model performance compared to FP 16 baseline.
Another work on large language models~\cite{liu_llm_fp4_2023} explored power precision, which reported only 0.9\% average accuracy drop for FP4 weight and FP8 activation for Llama2-7B~\cite{touvron2023llama} model.
That work further explored the usefulness of non-trivial precision such as FP6, reporting only 0.002\% accuracy drop on Vision Transformer~\cite{dosovitskiy2020image} for ImageNet~\cite{deng2009imagenet}.
However, although such works provide some computational performance benefits via software optimization (e.g., SIMT parallel dequantization~\cite{xia2024fp6}, current hardware unfortunately requires casting low-precision FP data to the supported precision FP data type (e.g., FP6 -> FP8).

Another trend is the use of various formats within the same bit-width, which assigns different numbers of bits to the exponent and mantissa (e.g., e4m3 and e5m2 for FP8~\cite{micikevicius2022fp8}, BF16~\cite{burgess2019bfloat16} and FP16~\cite{IEEE754}).
In addition, a recent work explored a completely flexible format within the precision to provide adaptivity to various dynamic range and precision requirements within a model~\cite{liu_llm_fp4_2023}.
Maintaining the hardware efficiency for such variants is yet another challenge to the hardware since most compute units in current hardware are specialized for specific data types (e.g., TPUv4 supports BF16 and INT8~\cite{jouppi2023tpu}), following the principle of specialization for accelerator design. 
Multiplications and accumulations in non-power-of-two precisions are dominant in aforementioned works. As activations remain in FP16, weights are usually stored in low-precision data types (e.g., FP6 for FP6-LLM~\cite{xia2024fp6}). Instead of dequantizing the weights to FP16, which is at the cost of hardware support and latency, weights are multiplied with FP16 activations directly to generate FP20 results assuming FP6 is e2m3 format, and accumulators thus perform FP20 accumulations.
We also claim that non-standard precisions will be more widely adopted once flexible hardware like \accelerator emerges, since it allows more fine-grained quantization design space (trade-off between accuracy/perplexity and efficiency) exploration than power-of-two precisions (e.g., to reduce precision from FP8, we need to jump to FP4, ignoring 5,6, and 7 bit options.).
Unfortunately, current hardware lacks flexible precision/format supports, as discussed in~\autoref{sec:intro}.
We discuss the details next.

%


%



\subsection{Hardware Support for FP Precision and Formats}
\label{subsec:hardware_support}

While there are many variants of floating-point arbitrary precision and formats in use at the algorithm level, current hardware architectures support a preset of floating-point precision and formats.
For example, NVIDIA H100 GPU includes separate FP8 and FP16 compute units~\cite{nvidia_h100}, and they are used exclusively; when FP16 operations are running, FP8 units are idle.
Such hardware resource underutilization is one of the main challenges for exploiting benefits from new data precision and formats, as illustrated in~\autoref{fig:Motivation} (c) (Challenge 1).

As another challenge, existing arithmetic units only support power-of-two precision, which requires up-casting of non-trivial formats to power-of-two precision, as illustrated in~\autoref{fig:Motivation} (c) (Challenge 2).
Such up-casting requires padding zeros (or ones depending on the sign) to the data, which results in fine-grained hardware resource underutilization (e.g., FP8 units computing FP6 operations neutralize arithmetic units and memory for two bits).
Such a limitation is a major challenge for advanced FP quantization such as FP6-LLM~\cite{xia2024fp6} and GPTQ~\cite{frantar2022gptq}, which limits the computational performance and energy benefits from non-trivial low precision formats such as FP5 and hybrid number formats (e.g., FP16 x INT4).
For example, GPTQ~\cite{frantar2022gptq} explicitly states that their method does not provide speedups due to the lack of hardware support for mixed-precision operands (e.g. FP16 x INT4) on mainstream architecture.
Likewise, current hardware architecture constraints algorithmic innovations in various quantization attempts for language models, which strongly motivated this work to develop an effective fully flexible FP precision/format accelerator, \accelerator.

As a potential solution, bit-serial accelerators such as Stripes~\cite{judd2016stripes} provide full flexibility by temporally processing each bit (i.e., serial manner) of operands.
Many bit-serial accelerators employ near- or in-memory computing due to the memory microarchitecture is friendly to bit level processing (e.g., in-cache accelerator~\cite{eckert2018neural}).
Although such solutions provide excellent efficiency, they face a major challenge for heavy workloads such as LLMs due to its limitation originating from memory technology (e.g., scale and clock frequency; DRISA~\cite{drisa}'s adder runs at \~1 MHz due to long tRC).
In addition, the computation in bit-serial accelerators is slowed down at the scale of bit-width due to its bit-serial processing nature, which is a key challenge for handling recent heavy workloads such as LLMs.
%
%
Such limitations of existing hardware architectures for large language models today with various FP precision and formats motivate us to design a new hardware architecture satisfying the following requirements:

\begin{itemize}
    {\item Efficiently supporting arbitrary FP precision/formats; both in compute units and memory}
    {\item A bit-parallel architecture to provide desired performance for compute-heavy large language models}
    {\item A scalable architecture for large language models}
\end{itemize}

Therefore, we develop \accelerator, a new accelerator architecture supporting fully flexible FP/INT precision/format, which satisfies all the requirements listed above.
We discuss the flexible compute unit architecture and introduce \accelerator deploying the flexible compute unit next.






\section{Flexible Precision Arithmetic Units in \accelerator}
\label{sec:FPALU}

%
To provide desired performance for recent compute-heavy language models with various FP/INT precision and formats, we aim to (1) design a bit-parallel architecture that performs arithmetic operations in high performance and (2) enable full flexibility for the ALU without underutilizing compute logic for arbitrary FP precision and formats.
%
%
Accordingly, we develop a new floating point multiplication unit based on our new architecture, flexible bit reduction tree (\ourtree), which allows parallel multiplication of arbitrary-length mantissa without compute unit underutilization.
\ourtree extends the the Augmented Reduction Tree (ART) in MAERI~\cite{kwon2018maeri} for the bit-level flexibility.
Our key novelty beyond the ART is (1) bit-level reconfigurability, unlike that in ART's is data element level, (2) extended node operations (flexible bit shifting, concatenation, and addition, while ART only performs additions), and (3) our carefully designed efficient control modes for complex node operations.

\insertFigureShrink{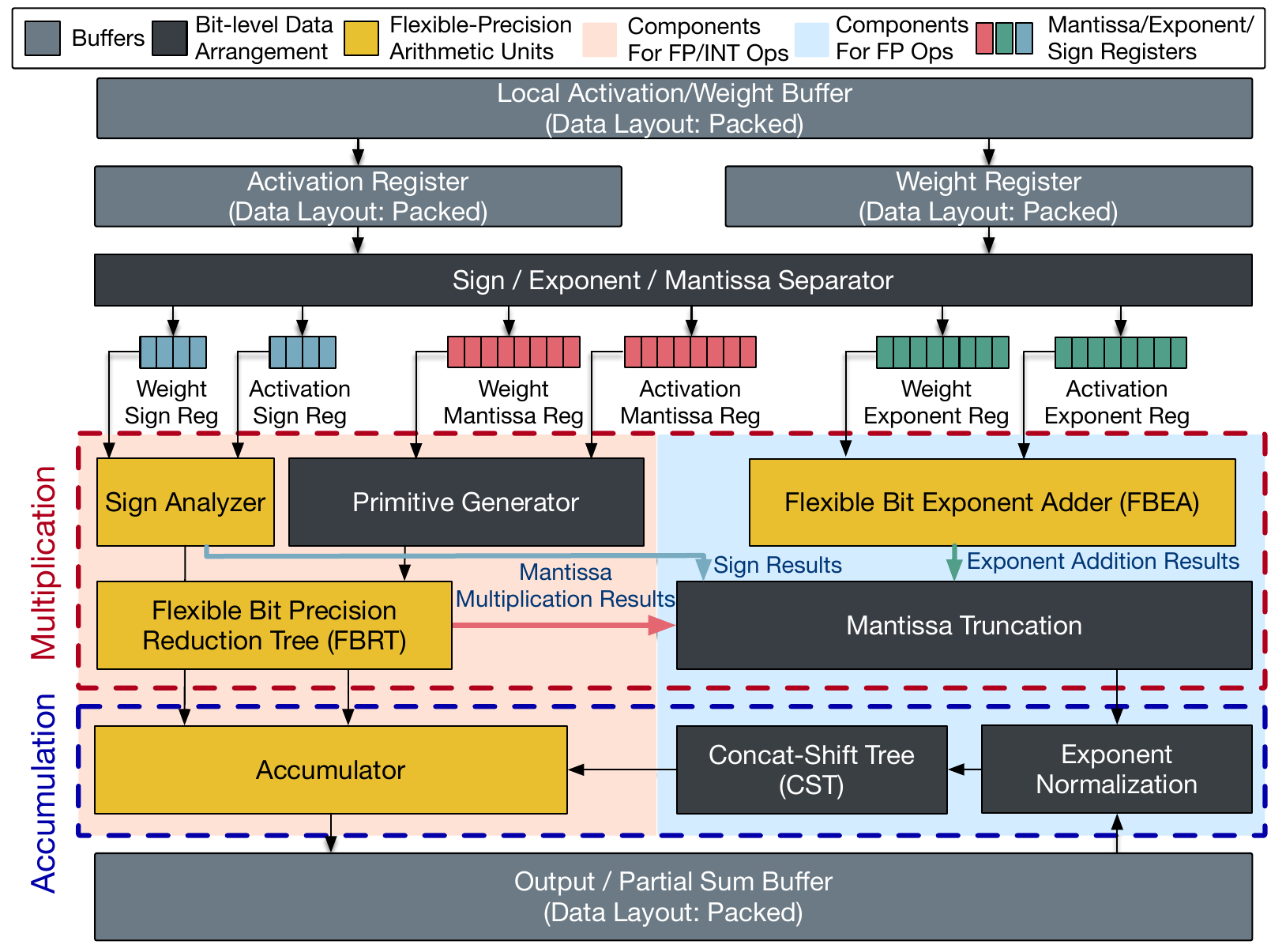}{\accelerator processing element (PE) architecture}{1}

\ourtree architecture enables aggregation of any bit-level multiplication primitives (i.e., cross-product AND results of each operand bit), which is the key enabler for flexible precision.
In addition to the \ourtree-based flexible multiplication, \accelerator's processing element (PE), illustrated in~\autoref{fig:PE_Architecture}, contains modules for flexible-bit exponent addition, truncation and shifting of bits, exponent normalization, and temporal accumulation of partial outputs (temporal accumulation is activated if required by dataflow). 
%
We provide an overview of the PE architecture and discuss each component next.
%

%
%


\begin{table}[t]
\centering
\scriptsize
\caption{Data annotation and PE design parameters choices}
\vspace{-2mm}
\resizebox{\columnwidth}{!}{
\begin{tabular}{| l | l | | l | l | r |}
\hline
\textbf{Annotation} & \textbf{Description} & \textbf{Parameter} & \textbf{Description} & \textbf{Val.} \\
\hline
$BW_M(A)$ & Activation mantissa bit width & $reg\_width$ & Weight/act register bit width & 24\\
$BW_M(W)$ & Weight mantissa bit width & $R_M$ & Mantissa register bit width & 12 \\
$BW_E(A)$ & Activation exponent bit width & $R_E$ & Exponent register bit width & 12\\
$BW_E(W)$ & Weight exponent bit width & $R_S$ & Sign register bit width & 12\\
$P(A)$ & Data precision of activation values & $L_{\text{prim}}$ & Primitive generator bit width & 144\\
$P(W)$ & Data precision of weight values & $L_{Add}$  & FBEA bit width & 144\\
 & & $L_{Acc}$ & Accumulator bit width & 144 \\
 & & $L_{CST}$ & Concat-shift Tree bit width & 144 \\
\hline
\end{tabular}
}
\label{tab:pe_params}
\vspace{-3mm}
\end{table}

\insertWideFigureShrink{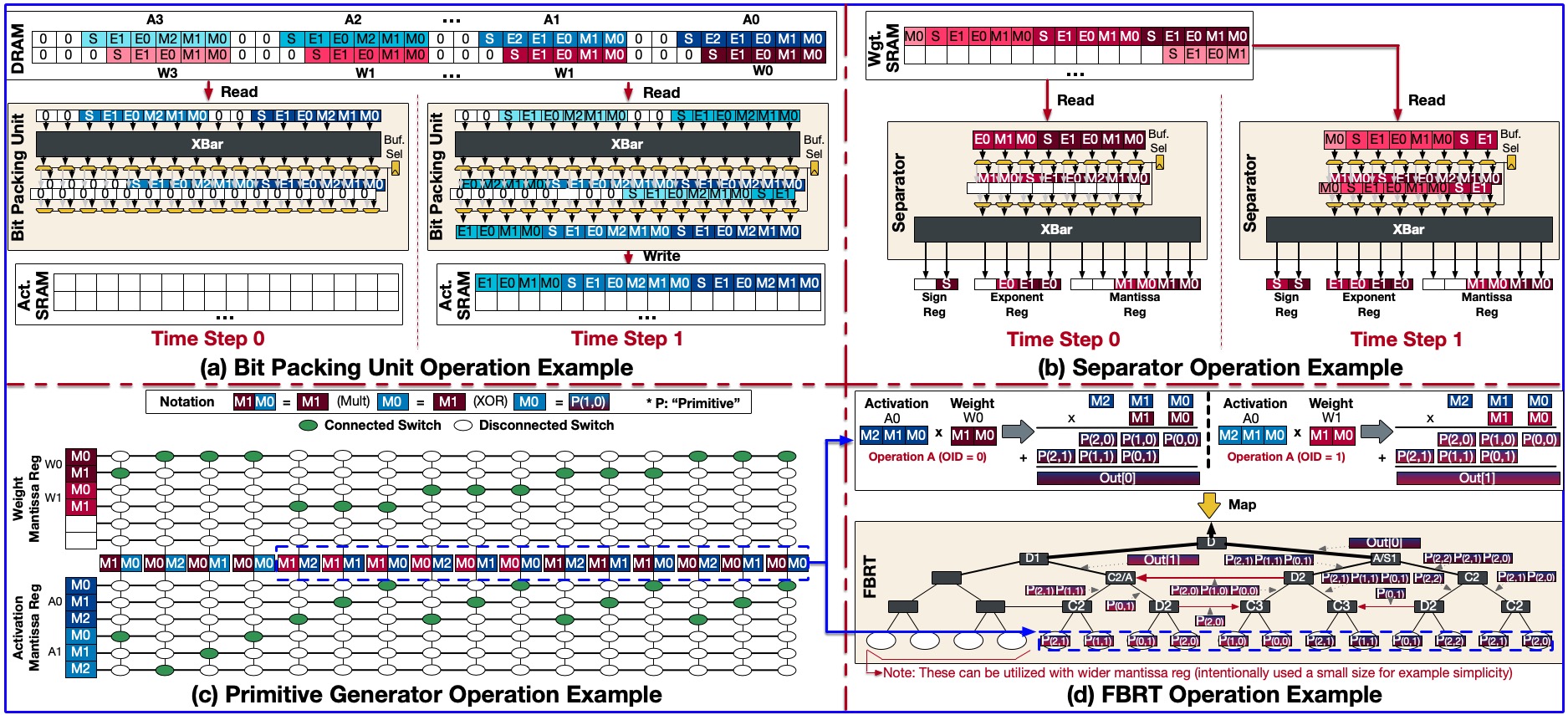}{A walk-through operation example of FP6 activation and FP5 weight from Bit Packing unit to FBRT. For Separator in (b), we only show the weights since its operation is the same for activation.}{1}


\subsection{PE Architecture Overview}
\label{subsec:arch_overview}

The Processing Element in~\nickname supports both FP and INT multiplication and addition, as illustrated in \autoref{fig:PE_Architecture}. We categorize hardware modules into two groups by their usage: One group dedicated to FP multiplication (red dotted box in~\autoref{fig:PE_Architecture}) and the other group for FP accumulation (blue dotted box in~\autoref{fig:PE_Architecture}).
The PE architecture has design time configurable design parameters as listed in~\autoref{tab:pe_params}. One key parameter that significantly affects the performance and cost is $reg\_width$. $reg\_width$ refers to the bit width of activation and weight registers and dictates the amount of parallelism offered by the PE datapath. Note that \accelerator's PE can also handle integer data. We highlight hardware components used for both integer/FP data and only for FP in ~\autoref{fig:PE_Architecture}.
For integer arithmetic, we bypass FP-dedicated hardware modules, which include flexible bit exponent adder and exponent normalization. We discuss details of each component following the datapath flow from the input operands to the output next.

\subsection{Sign / Exponent / Mantissa Separator}
\label{sec:separator}

For FP data, we need to separate mantissa, exponent, and sign bits for further processing. Although the separation is simple for standard FP formats via static bit split, it involves two major challenges for flexible precision FP data types:

\squishlist
    {\item The bit split point for sign, exponent, and mantissa can be any position since the input can be any FP data format.}
    {\item We need to process multiple FP data packed back-to-back without padding.}
\squishend 

%

The implication of the challenges is that we need a flexible bit routing mechanism to handle arbitrary cases. Our design strategy is enabling flexible bit routability using small crossbars and restricting the width of processed data at one time (i.e., $register\_width$) to constrain the hardware costs for crossbars. For example, if we adopt $reg\_width = 24$ and $R_M = 12$ and $R_E = 12$, we need $24 \times 12$ crossbars to provide full flexibility. Likewise, the size of crossbar switches are determined based on $R_M(A)$ and $R_E(A))$. To clarify the operation in the separator module, we provide an example in~\autoref{fig:Walkthrough} (b).
\\
\betterparagraph{Control Signal Generation} Generating correct route information to separate sign, exponent, and mantissa bits is crucial for the flexible precision separator. Using the data type parameters listed in~\autoref{tab:pe_params}, we describe the control logic as follows:

\begin{lstlisting}[language=Python,caption={A pseudo code for Separator control logic},captionpos=b, label=lst:separator]
mantissa_idx = 0;
exponent_idx = 0;
sign_idx = 0;

for i in range(reg_width):
  act_id = reg_width / P(A)
  act_bitid = reg_width % P(A)
  if act_bitid == 0 :
    act_sign_reg[sign_idx] = act_reg[i]
    sign_idx++
  elif act_bitid < 1 + BW_E(A):
    act_exponent_reg[exponent_idx] = act_reg[i]
    exponent_idx++
  else
    act_mantissa_reg[mantissa_idx] = act_reg[i]
    mantissa_idx++ 
\end{lstlisting}

~\autoref{lst:separator} describes the Primitive Generator control logic using a Python-style pseudo code for activation. In~\autoref{lst:separator}, act\_sign\_reg, act\_mantissa\_reg, and act\_exponent\_reg refer to the corresponding registers in~\autoref{fig:PE_Architecture} (blue, red, and green registers). The same logic is applied to weight as well. Note that control signals are identical across the PEs for the same layer. Therefore, we broadcast the control signals to all the PEs.

\subsection{Primitive Generator}
\label{sec:prim_gen}

We can observe that the multiplication process consists of (1) computing AND for each operand bit pair and (2) bit shifting and addition. \accelerator's PE also follows those steps. For the first step, we design a hardware module named "Primitive Generator." We term the individual AND results for each activation-weight bit pair as "Primitive." As notation, we refer to the primitive generated between the j-th bit of activation A and i-th bit of weight as $P_A(i,j)$, which is defined as $P_A(i,j) = A_i \cdot W_j$ where $\cdot$ refers to the bit-wise AND operator.


%

As inputs, Primitive Generator receives a set of activation and weight mantissa stored in activation and weight mantissa registers (red registers in~\autoref{fig:PE_Architecture}), respectively. Primitive Generator parses the bits in the activation and weight mantissa registers to generate primitives in the order and layout consumed by \ourtree, which will be discussed in~\autoref{sec:FBRT}. Since (1) the mantissa bit-widths can be different for both inputs and activations and (2) we want the primitives of each operation to be alongside each other for FBRT, different patterns can occur when parsing the bits. For the implementation, we take the same approach as the Separator discussed in~\autoref{sec:separator}: utilize small crossbars and keep a reasonable width to constrain the hardware costs while providing performance benefits.

~\autoref{fig:Walkthrough} (c) illustrates an operation example of Primitive Generator for $BW_M(A) = 3$ and $BW_M(W) = 2$. To compute the multiplication given in~\autoref{fig:Walkthrough} (c), Primitive Generator needs to generate the cross-product of each activation and weight bit. We use two crossbars for weights and activations and match the corresponding operands for each primitive. Note that we generate the primitive in the ascending order of weight/activation bit indices in a packed format, as shown in ~\autoref{fig:Walkthrough} (c). Maintaining the order is crucial for the operation of Flexible-Bit Reduction Tree (FBRT), which is the key enabler for the flexible precision/format multiplication.
Note that we handle the implicit-1 bit in another module to prevent the exponential enlargement of the crossbar and \ourtree size, which will be discussed in~\autoref{sec:FBRT}. 

\betterparagraph{Control Logic} We describe the control logic as follows:

\begin{lstlisting}[language=Python,caption={A pseudo code for Primitive Generator control logic},captionpos=b, label=lst:prim_gen]
num_prims = BW_M(A) * BW_M(W)
num_acts = R_M / BW_M(A)
num_wgts = R_M / BW_M(W)

for i in range(L_prim):
  output_id = i / num_prims
  act_id = i / num_prims
  wgt_id = i / (num_acts * num_prims)
  if act_id < num_acts and wgt_id < num_wgts:
      prim_reg[i] = act_mantissa_reg[act_id] & wgt_mantissa_reg[wgt_id]

\end{lstlisting}
\vspace{-3mm}

In~\autoref{lst:prim_gen}, prim\_reg refers to a register where the output of primitive generation is stored. The length of prim\_reg is $L_{prim}$, defined in~\autoref{tab:pe_params}.

\subsection{FBRT: A Flexible-Bit Reduction Tree}
\label{sec:FBRT}


\ourtree is the key enabler for the flexible-precision bit-parallel multiplication. \ourtree approaches the multiplication operation as a spatial shift-add operation and aggregates primitive bits in a tree architecture inspired by inspired by Augmented Reduction Tree (ART) of MAERI~\cite{kwon2018maeri}. Our contribution beyond the ART is (1) the fine-grained (bit-level) reconfigurability, (2) complex node operations (add, concat, shifting, and their combinations), and (3) control modes that coordinate complex node operations. 
What we mainly adopt from ART is the fat-tree-like structure and additional links, which are links between neighboring nodes in the same level that do not share the same parent node. \ourtree expects the primitives are sorted in terms of activation and weight indices, as illustrated in~\autoref{fig:Walkthrough} (d), which is managed by Primitive Generator. The primitives traverse pre-configured tree from leaf nodes to simultaneously generate multiple multiplication results. The tree nodes have four operating modes with variants within the mode to implement the core functionality, which we discuss next.

\insertFigureShrinkCustomMargin{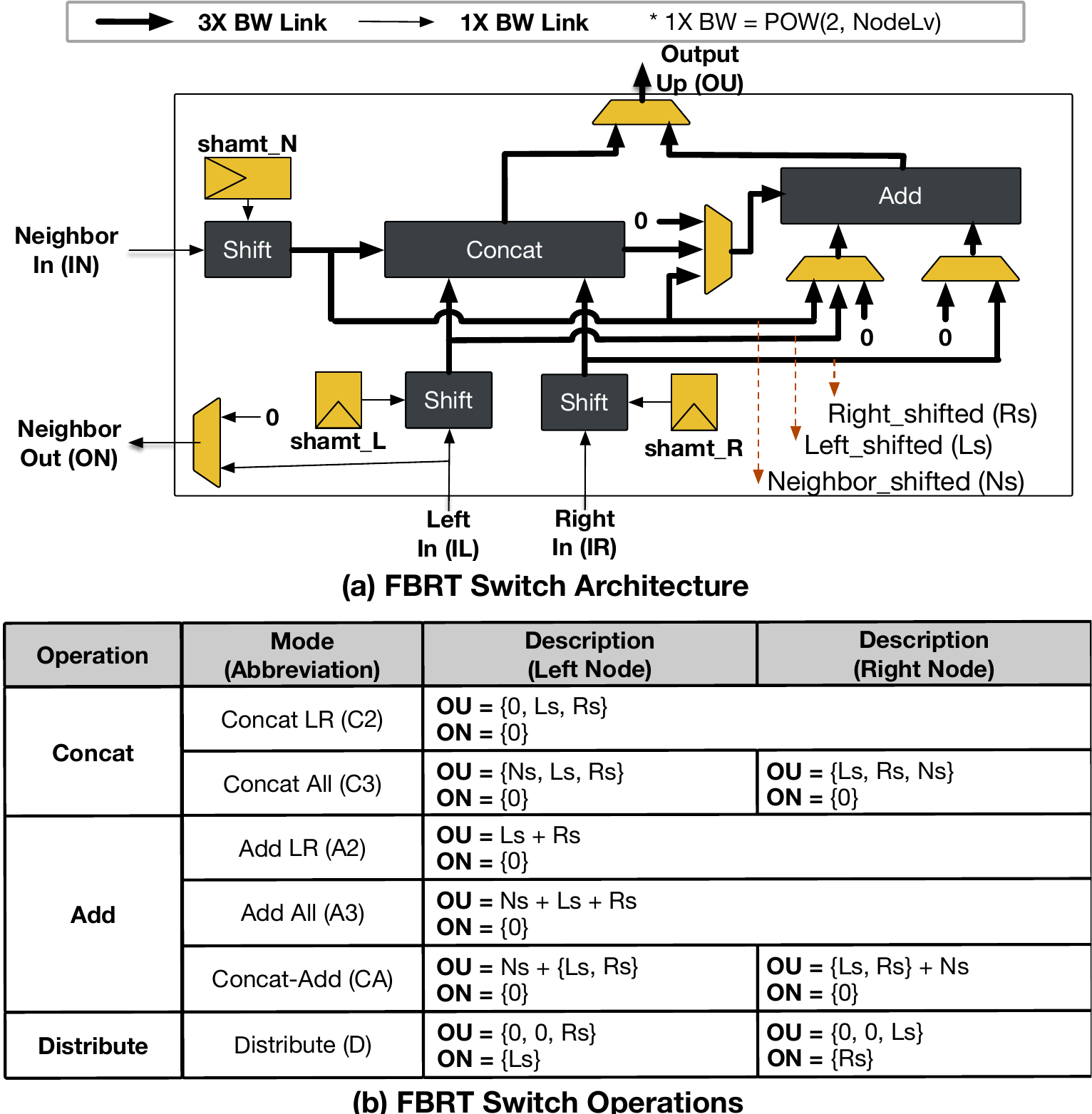}{Microarchitecture switch and possible modes overview. Left and Right nodes refer to the tree node with neighbor link on the left and right side, respectively. The brackets "\{" and "\}" refer to concatenation.}{1}{-7}{-5}

\betterparagraph{Switch Modes} We summarize the switch modes of \ourtree in~\autoref{fig:Mode_Control}. As shown in the table, \ourtree's node switches support four modes with variants depending on the position of the node (left or right to the additional link) and the number of operands a switch handles. The operation of the switches consists of routing, concatenation, shifting, and addition, as shown in the operation description in~\autoref{fig:Mode_Control}. To understand how those switches are used to enable bit-parallel flexible precision multiplication, we discuss an operation example next.

\betterparagraph{Operation Example} ~\autoref{fig:Walkthrough} (d) shows an example operation of \ourtree on the multiplication between FP6 activations and FP5 weights
In the example, red arrows refer to the use of additional links, which is crucial for the bit level flexibility.
Various switch mode labels on each tree node show how the complex node operations are combined to enable the flexible bit arithmetic.
For example, Out[1] is generated by concatenating two incoming primitives from dependent nodes, left-shifting it by one, and adding it with the other incoming primitive from the additional link (\{P(2,0), P(1,0), P(0,0\} + ((concat ((\{ P(2,1),P(1,1)\}), P(1,0))) << 1).
Likewise, primitives are progressively concatenated and added after aligning the bit positions using shifting.
Although the control logic on switches handles complex operations, the logic is controlled by a simple finite state machine (FSM) using compiler-generated control signals, which we discuss next.


\betterparagraph{Control Logic} Since the data precision and format are expected to be consistent within each layer, we can utilize a compiler to statically generate the control signals. We describe the algorithm below in a Python-style pseudo code:

\begin{lstlisting}[language=Python,caption={Pseudo code of FBRT control signal generation},captionpos=b, label=lst:fbrt_control]
// Inputs
//oids: array - Output IDs for each bit position in the primitive reg.
//sids: array - Segment IDs for each bit position in the primitive reg.

curr_oids = oids
curr_sids = sids

for i in range(1,num_levels):
  oids = odis.append(curr_oids)
  sids = sids.append(curr_sids)
  
  num_nodes_this_lv = FBRT_width / pow(2,i)
  for j in range(num_nodes_this_lv,0,-1):
   if has_additional_links(i,j):
    if oids[i-1][j*2]+1 == oids[i-1][j*2]:
     if sids[i-1][j*2]+1 == sids[i-1][j*2]:
      if oids[i-1][j*2]+1 == get_N_oids[i][j]:
        switch_mode[i][j] = C3
      else:
        switch_mode[i][j] = C2   
     else: // if children SIDs do not match 
      if oids[i-1][j*2]+1 == get_N_oids[i][j]:
       if sids[i-1][j*2]+1 == get_N_sids[i][j]:
         switch_mode[i][j] = Concat_Add
       else:
         switch_mode[i][j] = A3
      else:
        switch_mode[i][j] = A2
    else: // if children OIDs do not match
       switch_mode[i][j] = D

\end{lstlisting}
\vspace{-2mm}

The control signal generation algorithm above simulates the operation in a compiler from the leaf nodes to the top (root node), filling switch modes for each switch in a level from right to left. To determine the switch mode, the algorithm utilizes the operation ID (OID), which refers to the ID of different multiplications, and segment ID, which refers to the ID of rows of intermediate results shown in~\autoref{fig:Implicit_1}. Note that a compiler handles the control signal generation, which does not incur runtime overheads.

\betterparagraph{Optimization for the implicit 1} FP formats assume implicit 1 in the mantissa for the efficiency of representation. One way to directly support implicit 1 is by adding 1s when we generate primitives. However, this approach leads to a significantly larger number of primitives. For example, if we multiply two mantissa with 2-bit and 3-bit representations, we need 2 $\times$ 3 = 6 primitives. However, if we include implicit 1s, we need (2+1) $\times$ (3+1) = 12 primitives, which is double).
To avoid excessive overhead from implicit ones, we utilize its key characteristics: the value is always one. \autoref{fig:Implicit_1} presents our approach. We first generate intermediate outputs without implicit ones using \ourtree. After that, in step 1, we exploit the property that left-most values of segments are original weight bits (e.g., W0 and W1 appear in the left-most position of segments from the top in ~\autoref{fig:Implicit_1} step 1). That is, we can shift the original weight and add it to the results of \ourtree, P\_FBRT. Afterwards, we take the same approach for the activation, whose original values are listed in the bottom segment. Note that the left-most bit of the bottom segment is always 1 since it is the primitive that corresponds to the implicit 1s of activation and weight.


\insertFigure{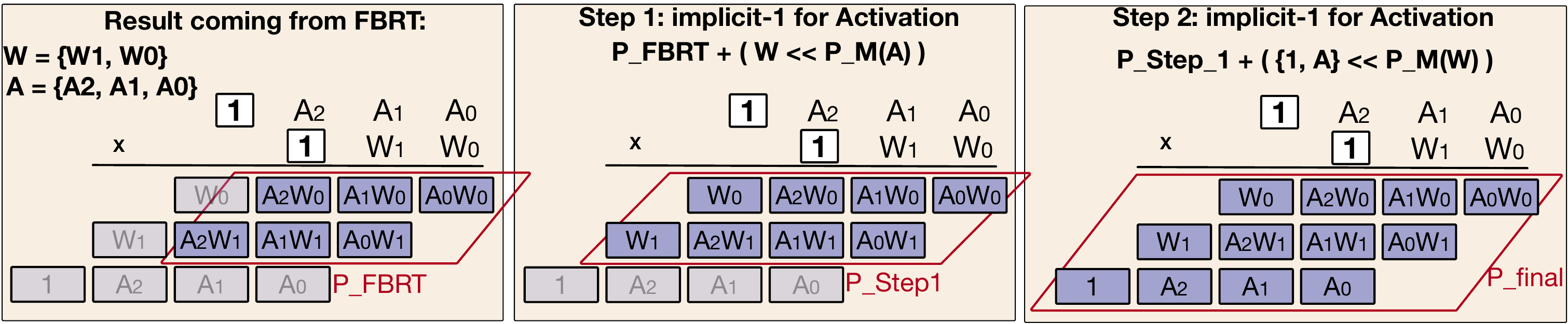}{How \ourtree handles implicit 1 for mantissa multiplication. P\_FBRT refers to the intermediate results generated in FBRT.\vspace{-2mm}}
\subsection{Flexible Bit Exponent Adder (FBEA)}
\label{sec:Exp_Adder}

To implement a flexible adder for exponents, we design a segmentable carry-chain adder. This adder can be flexibly partitioned for handling many low-precision data or a few high-precision data. \autoref{fig:Exp_Adder} demonstrates an example of an 18-bit flexible adder operating for $P_E(A) = 3$ and $P_E(W) = 2$. Between each full adder, a multiplexer controls the carry chain. If a new operation starts, ctrl breaks the carry chain. This simple yet effective architecture provides the desired capability of exponent addition for any arbitrary precisions.

\betterparagraph{Control Logic} The control signal has the same width as the adder. The values of 0 and 1 indicate that the carry needs to propagate and stop, respectively. This indicates the boundary of additions. We describe the control logic as follows:

\begin{lstlisting}[language=Python,caption={Pseudo code of FBEA control logic},captionpos=b, label=lst:adder_control]
add_width = max(BW_E(A), BW_E(W)

for i in range(L_add):
  if (i + 1) % add_width == 0:
    FBEA_control[i] = 1
  else:
    FBEA_control[i] = 0

\end{lstlisting}
\vspace{-4mm}


\insertFigureShrinkCustomMargin{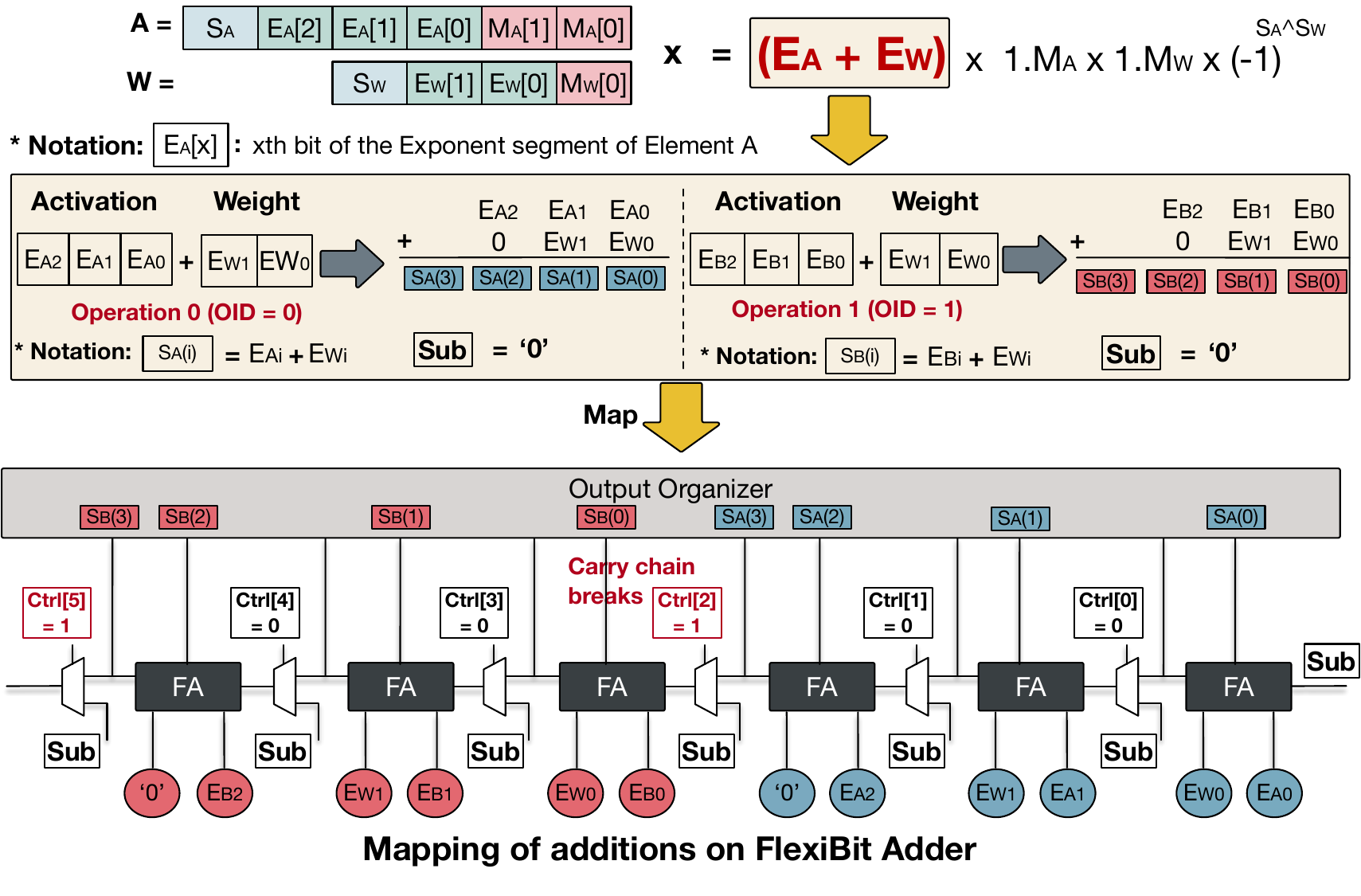}{Flexible bit exponent adder microarchitecture}{1}{-7}{-5}

\subsection{Exponent Normalization Unit (ENU)}
\label{sec:Exponent Normalization}

Exponent normalization unit (ENU) handles the exponent alignment step described in the floating point addition process in~\autoref{subsec:fp_arithmetic}. ENU first parses incoming bit-packed exponents using a similar parsing algorithm used for Primitive Generator discussed in~\autoref{sec:prim_gen}. It subtracts corresponding exponents and identifies the shift amount. 


\subsection{Concat-Shift Tree (CST)}
\label{sec:shift_tree}

Using the shift amount results generated by ENU, concat-shift tree shifts one of the mantissas to align the scale of two mantissas. The policy for selecting shifted exponent (e.g., shift the higher exponent to the lower one) is user-configurable.

~\autoref{fig:Shift_tree} (a) illustrates an example operation of CST with three-bit mantissas. The difference in the exponent computed by ENU is used to determine the shift amount of each level. The control logic generation follows a similar approach as \ourtree, tracking the ID of individual mantissa and concatenating them if values from left and right children belong to the same mantissa ID. If a value from the Additional link belongs to the same mantissa as well, a tree node switch performs three-way concatenation. Note that the shift amount computed from ENU for individual mantissa is applied when performing concat-shift operation.

\insertFigureShrink{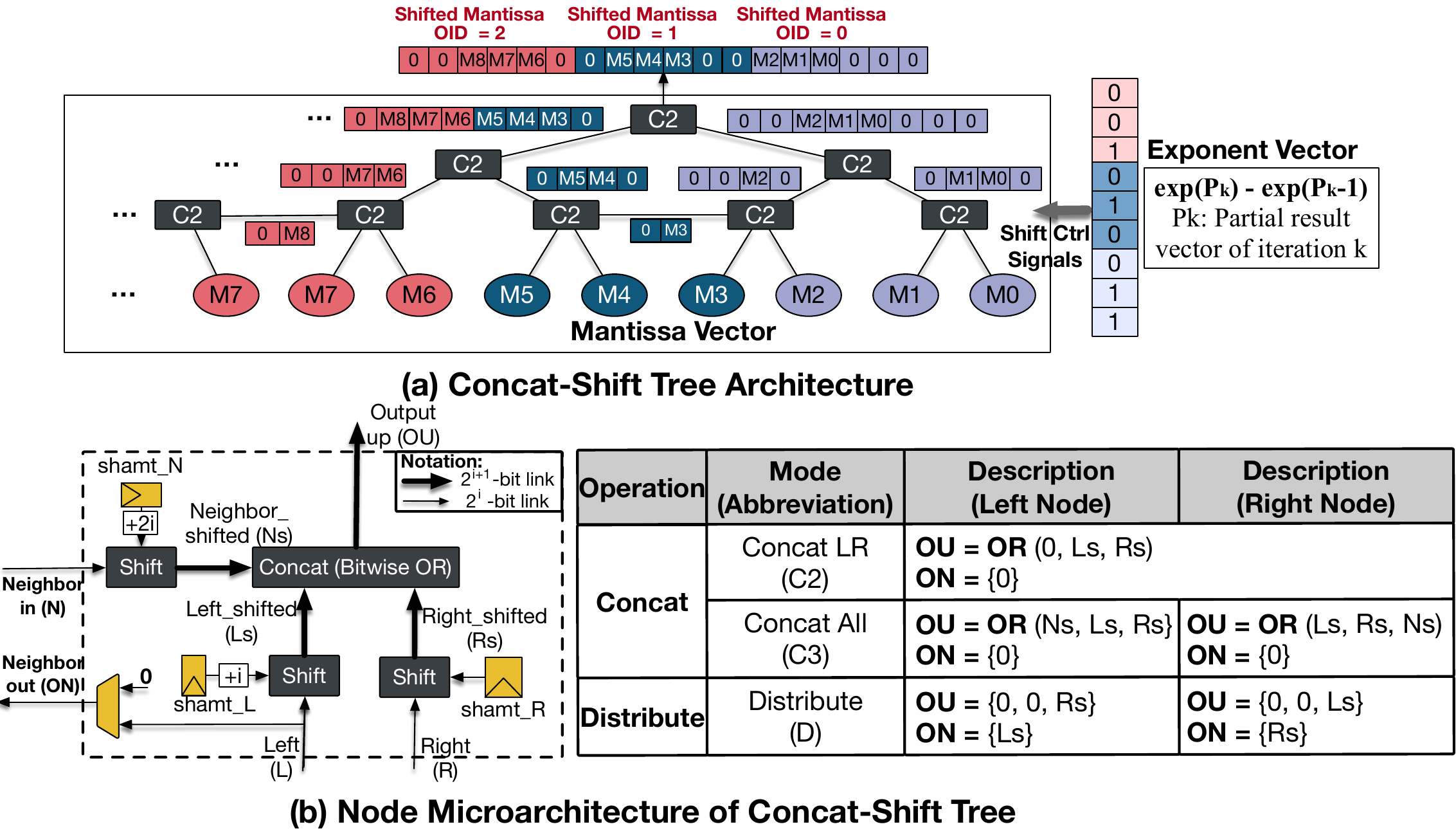}{(a) The Concat-Shift Tree (CST) architecture and an example operation on 3-bit mantissas. (b) Node microarchitecture and modes in CST.}{1}

\subsection{Accumulation and Normalization Unit (ANU)}
\label{sec:accumulator_normalizer}

Based on the results from CST, the Accumulation and Normalization Unit (ANU) adds incoming partial outputs, normalizes the exponent based on the results of addition, and shifts the mantissa considering (1) the implicit 1, (2) newly normalized exponent, and (3) target output precision. We adopt the FBEA unit and its control mechanism to implement the core functionalities of ANU.

\subsection{Microscaling(MX)-Format support} 

As described in~\autoref{subsec:fp_arithmetic}, the main difference between the MX arithmetic from standard FP arithmetic is in the shared scaling factor. Therefore, to support MX formats, we embed two dedicated registers in each PE for the scaling factors. We apply the scaling factors when we determine the final results and report the output in the MX format. This process does not incur noticeable latency due to its simple operation and frequency (the scaling factor remains the same for many data elements).
Based on the PE architecture discussed in this section, we present how we organize them into a full accelerator next.

\section{\nickname: Fully-flexible Precision Accelerator}
\label{sec:architecture}

\insertFigureShrink{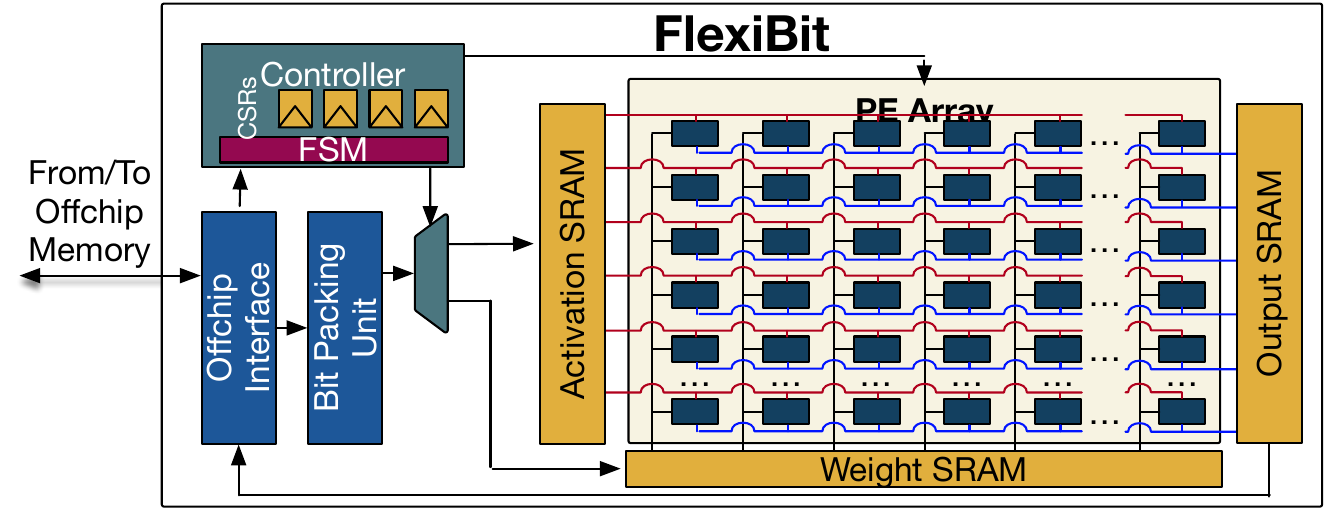}{Overview of \nickname accelerator}{0.8}
\autoref{fig:Accelerator} depicts an overview of our fully-flexible precision/format bit-parallel accelerator, \accelerator. \accelerator follows the general architecture of AI accelerators at the high level but it contains new architectures for flexible precision and format in the PE and bit-packing and -unpacking unit. \accelerator's control signals are generated by a compiler since they are deterministic when the precisions are determined. The control signal generation algorithm also can be implemented in a hardware controller, which can be used for dynamic control of the hardware. However, as the granularity of precision/format change is coarse-grained in current practice (layer/operator or model level), we utilize compiler and store the control signals in the control and state registers (CSRs) as showin in ~\autoref{fig:Accelerator}.
With this approach, the controller's area overhead is negligible (~0.2\% of the total accelerator area). Note that the control signals can be broadcasted to PEs if the precision changes in the layer or model granularity, which minimizes the reconfiguration cost (less than 100 cycles; note that each GEMM consumes millions-billions of cycles). 

We focus on accelerator-level features with less PE array in this section. We first discuss $BitPacking$ and $Unpacking$ units to show how we store data in on-chip memories in a bit-packed format and how we unpack the packed data before sending the data to an off-chip memory.


\subsection{Bit Packing and Unpacking Unit (BPU)} 
Storing and accessing such non-power-of-two precision data introduces a major challenge to the memory and network-on-chip. One simple approach is to pad zeros with the remaining bits (e.g., store FP6 data into an 8-bit memory block with padding). Dispite simplified control logic, this approach is not ideal for on-chip memory efficiency. Therefore, we implement a crossbar-based bit-packing unit (BPU) as illustrated in~\autoref{fig:Walkthrough} (a), to support condensed bit-packed memory layout in the accelerator. The BPU is activated if the host memory utilizes the zero-padded memory layout. The incoming data is 64 bits with zero-padded data when the offchip interface is 64-bit wide. The controller utilizes precision information (i.e. 6 in this example) and a $start\_idx$ register. 
Each useful bit in the $i$-th position of the input is mapped to $j$'th position of the output. $j$ is calculated using this formula: $j = start\_idx + i - (\lfloor i/8\rfloor \times (8 - precision)$. \autoref{fig:Walkthrough} (a) demonstrates an example with FP6 activation. First six bits on the activation are mapped with the same index. 7th and 8th bits of input are masked and has no effect on the output vector. Input bits of 9 to 14 are mapped to output bits of 7 to 12.  In the next data arrival, $start\_idx = start\_idx + precision * 8$ is applied and the crossbar maps each input bit using the same method. A double buffering is considered to handle packing overflow properly and transferring the data to the SRAM. BPU is also responsible for propagating metadata to the accelerator controller, indicating the starting address, bit precision information, and number of data elements.

\noindent
\betterparagraph{Bit-packing vs Padding} Padded data formats are preferred in system software (OS, runtime, firmware, etc.) due to their data address alignment requirements, which is crucial for the functionality and efficiency of sub-components of system software such as memory allocator. However, padded data involves redundant zeros, which leads to inefficiency. By removing the redundancy, packed data layout reduces the on-chip traffic, memory efficiency, and compute unit utilization. We'll discuss the overhead in ~\autoref{subsubsection:area_analysis}.

\subsection{Dataflow}
\label{subsec:dataflow}

Although \accelerator PE requires an outer-product style GEMM computation for efficiency, the dataflow in the PE array level is fully flexible in \accelerator, as long as each compute tile fits into each PE's local memory.
The key enabler for the flexible dataflow is the two-dimensional bus-based NoC, which allows data transfer to any PE from each SRAM.
In the evaluation, we select the best dataflow style among weight- and output-stationary dataflows in the PE array level, setting the tile size to be the maximum possible for on-chip memory size.



\section{Evaluation}
\label{sec:evaluation}



\subsection{Evaluation Settings}

\betterparagraph{Baselines} We compare our work against state-of-the-art flexible precision bit-parallel accelerator architecture, Bit-Fusion \cite{sharma2018bit} and a fixed precision bit-parallel accelerator architecture, Tensor Core ~\cite{nvidia_tensorcore}-style systolic array.
Note that Bit-Fusion is originally designed for integer arithmetic; we extend Bit-fusion for floating-point to focus on modeling their novel architecture for bit precision flexibility.
We adopt the reported area and power consumption of Bit-fusion and scale the technology.
%
%
For a fair comparison, we conduct iso-PE analysis across all evaluated accelerators.
We apply weight-stationary dataflow style following the original implementations of baselines.

\betterparagraph{LLM Workloads} We use four language models: Bert-base-uncased~\cite{kenton2019bert}, Llama-2-7b~\cite{touvron2023llama}, Llama-270b~\cite{touvron2023llama}, and GPT-3~\cite{GPT3}.
We list the hyperparameters of the evaluated models in~\autoref{tab:Workload_spec}.

\begin{table}[t]
\centering
\scriptsize
\caption{\accelerator configurations in evaluations.}
\vspace{-3mm}
\resizebox{1.0\columnwidth}{!}{
\begin{tabular}{|c|c|c|c|c|c|}
\hline
\textbf{Accelerator} & \textbf{Mobile-A} & \textbf{Mobile-B} & \textbf{Cloud-A} & \textbf{Cloud-B} \\
\hline
\textbf{Num. of PEs} & 1K PE & 4K PE & 8K PE & 16K PE \\
\hline
\textbf {Reg\_Width (bits)}  & 24  & 24  & 24 & 24 \\
\hline
\textbf{Off-chip}  & 16  & 16  & 128 & 128 \\
\textbf{Bandwidth (GB/s)}  & (DRAM)  & (DRAM)  & (HBM) & (HBM) \\
\hline
\textbf{Weight}  & \multirow{2}{*}{2}  & \multirow{2}{*}{4}  & \multirow{2}{*}{16}  & \multirow{2}{*}{32} \\
\textbf{Global Buffer (MB)}  &   &   &   &  \\
\hline
\textbf{Act/Output}  & \multirow{2}{*}{1}  & \multirow{2}{*}{2}  & \multirow{2}{*}{8}  & \multirow{2}{*}{16} \\
\textbf{Global Buffer (MB)}  &    &   &   &  \\
\hline
\textbf{W/A NoC}  & \multirow{2}{*}{32}  & \multirow{2}{*}{64} & \multirow{2}{*}{128/64} & \multirow{2}{*}{128} \\
\textbf{Bandwidth (GB/s)} &  &  &  &  \\
\hline
\textbf{PE Array}  & \multirow{2}{*}{$32 \times 32$}  & \multirow{2}{*}{$64 \times 64$}  & \multirow{2}{*}{$128 \times 64$} & \multirow{2}{*}{$128 \times 128$} \\
\textbf{($X \times Y$)}  &   &   &   &  \\
\hline
\textbf{Local Buffer}  & \multirow{2}{*}{0.18}  & \multirow{2}{*}{0.18}  & \multirow{2}{*}{0.18} & \multirow{2}{*}{0.18} \\
\textbf{per PE (KB)}  &   &   &   &  \\
\hline
\end{tabular}
}
\label{tab:HW_spec}
\vspace{-10mm}
\end{table}

\betterparagraph{Hardware Configurations:} We simulate the accelerators in four different configurations: Mobile-A, Mobile-B, Cloud-A, and Cloud-B.
We list the details of the configurations in~\autoref{tab:HW_spec}.
We apply the default hardware design parameters listed in~\autoref{tab:pe_params}, which was selected after cost-benefit analysis.
We deploy weight- and output-staionary dataflow styles for our design.
We set the design parameters to the default values listed in~\autoref{tab:pe_params}, which are optimized after exploring performance and area trade-offs. 

The design parameters of Mobile-A and -B are selected based on recent works~\cite{liu2024mobilellm, wen2024autodroid}. We also include two cloud-scale accelerators whose scales are modeled after TPUv4~\cite{jouppi2023tpu}. We adopt off-chip bandwidth parameters from Ramulator~\cite{kim2015ramulator} and recent industry work~\cite{jun2017hbm} for performance evaluation.


\subsection{Methodology}

We implement our design in RTL using System Verilog and conduct Post-PnR analysis using NanGate 15nm Technology~\cite{martins2015open} and Cadence Innovus.
We implement a cycle-accurate simulator designed to model our RTL implementation. The simulator is validated against RTL simulation with 96\% accuracy on Bert-base and 99\% accuracy on Llama-2-7b, as shown in~\autoref{fig:perf_valid}. We utilize a state-of-the-art energy estimation framework, Accelergy~\cite{wu2019accelergy}, for evaluating energy, with the synthesis results of our RTL implementation. We estimate DRAM read/write energy based on previously reported data~\cite{dram_energy}.

\insertFigureShrinkCustomMargin{perf_valid}{Performance model validation against RTL simulation on attention layers of Bert-base and Llama-2-7b.}{1}{-7}{-4}




\begin{table}[t]
\centering
\scriptsize
\caption{Hyper parameters of the evaluated models}
\vspace{-3mm}
\resizebox{\columnwidth}{!}{
\begin{tabular}{|c|c|c|c|c|}
\hline
\multirow{2}{*}{\textbf{Model Name}}  & Input  & Number of    & Embedding   & Hidden  \\
  &  Sequence  &   Layers   &  Dimension  &  Dimension \\
\hline
\textbf{Bert-Base-uncased}  & 2048  & 12  & 768  & 3072 \\
\hline
\textbf{Llama-2-7b}  &  2048  & 32  & 4096  & 11008 \\
\hline
\textbf{Llama-2-70b}  &  2048  & 80  & 8192  & 28672 \\
\hline
\textbf{GPT-3}  &  2048  & 96  & 12288  & 49152 \\
\hline
\end{tabular}
}
\label{tab:Workload_spec}
\vspace{-5mm}
\end{table}

\subsection{Evaluation Results}
\label{subsec:results}


\insertWideFigureShrinkCustomMargin{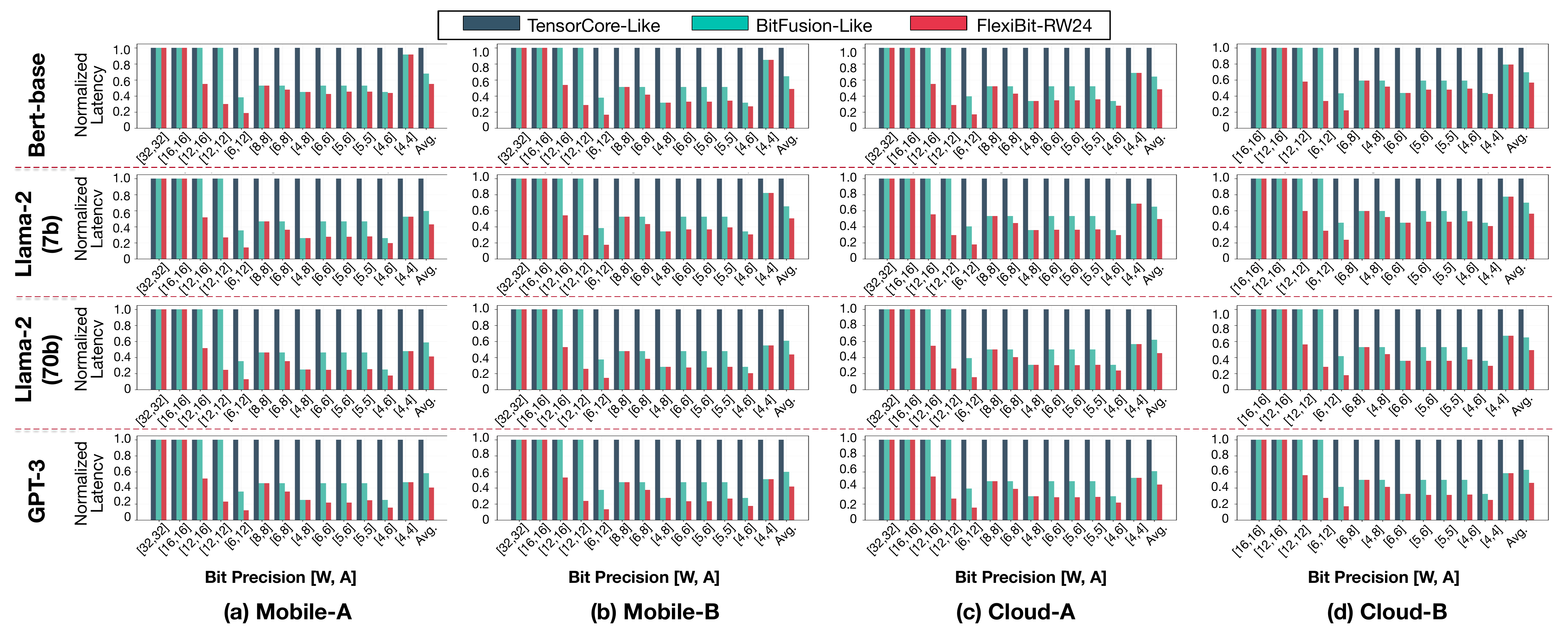}{Latency of Bert-Base, Llama-2-7b Llama-2-70b, and GPT-3 models on different precisions. Input sequence is 2048. Results are depicted for (a) Mobile-A configuration, (b) Mobile-B configuration, (c)Cloud-A configuration, and (d)Cloud-B configuration accelerator scales.}{1}{-7}{-4}

\subsubsection{Latency and Energy Comparison}
\label{sec:latency}


Our experiments demonstrate that \accelerator's performance varies for Output-Stationary (OS) and Weight-Stationary (WS) for different accelerator scales and workload models, which is aligned with previous works on dataflow~\cite{timeloop, kwon2019understanding}. The OS dataflow parallelizes M and N dimensions and reuse partial outputs K times. The WS dataflow parallelizes K and N dimensions and reuse weights M / TileSz(M) times. Their performance and efficiency varies depending on different data reuse counts, as formulated above, and degree of parallelization determined by the interplay between tensor shapes, tile sizes, and accelerator array sizes. In the evaluation, we leverage the dataflow flexibility of \accelerator and report results based on the best dataflow between WS and OS for each experiment. Note that the dataflow preference can be quickly determined using our performance model.

\autoref{fig:Latency} presents the latency evaluation results of \accelerator and baselines. We observe minor improvements for FP16-based models. However, as we deploy non-power-of-two precisions, \accelerator significantly outperforms all the baselines, providing 59\% and 31\% less latency compared to Tensor Core and Bit-Fusion, on average. The improvements mainly originate from the capability to support arbitrary precision with high hardware underutilization. Baselines need to up-cast non-power-of-two precision data to the nearest power-of-two precision data, which leads to significant redundant bit operations. Another source of improvement is our $BitPacking$ technique which fully utilizes the SRAM capacity, which reduces DRAM traffic. To better demonstrate the impact of $BitPacking$, we study end-to-end latency of~\accelerator with and without $BitPacking$ scheme, as presented in \autoref{fig:Latency_Bitpacked_Padded}. $BitPacking$ improve the latency by 26\%, on average, while the area and power overhead is negligible to the entire accelerator, as  \autoref{subsubsection:area_analysis} presents.

In terms of energy consumption,~\accelerator with weight-stationary dataflow provides considerably low energy, 66\% and 33\% lower than the Tensor Core and Bit-Fusion baselines, on average. The main source of energy saving is reduced on-chip traffic by utilizing packed memory layout. As a result, the energy saving is significant when non-power-of-two precision data is involved. 

\insertFigureShrinkCustomMargin{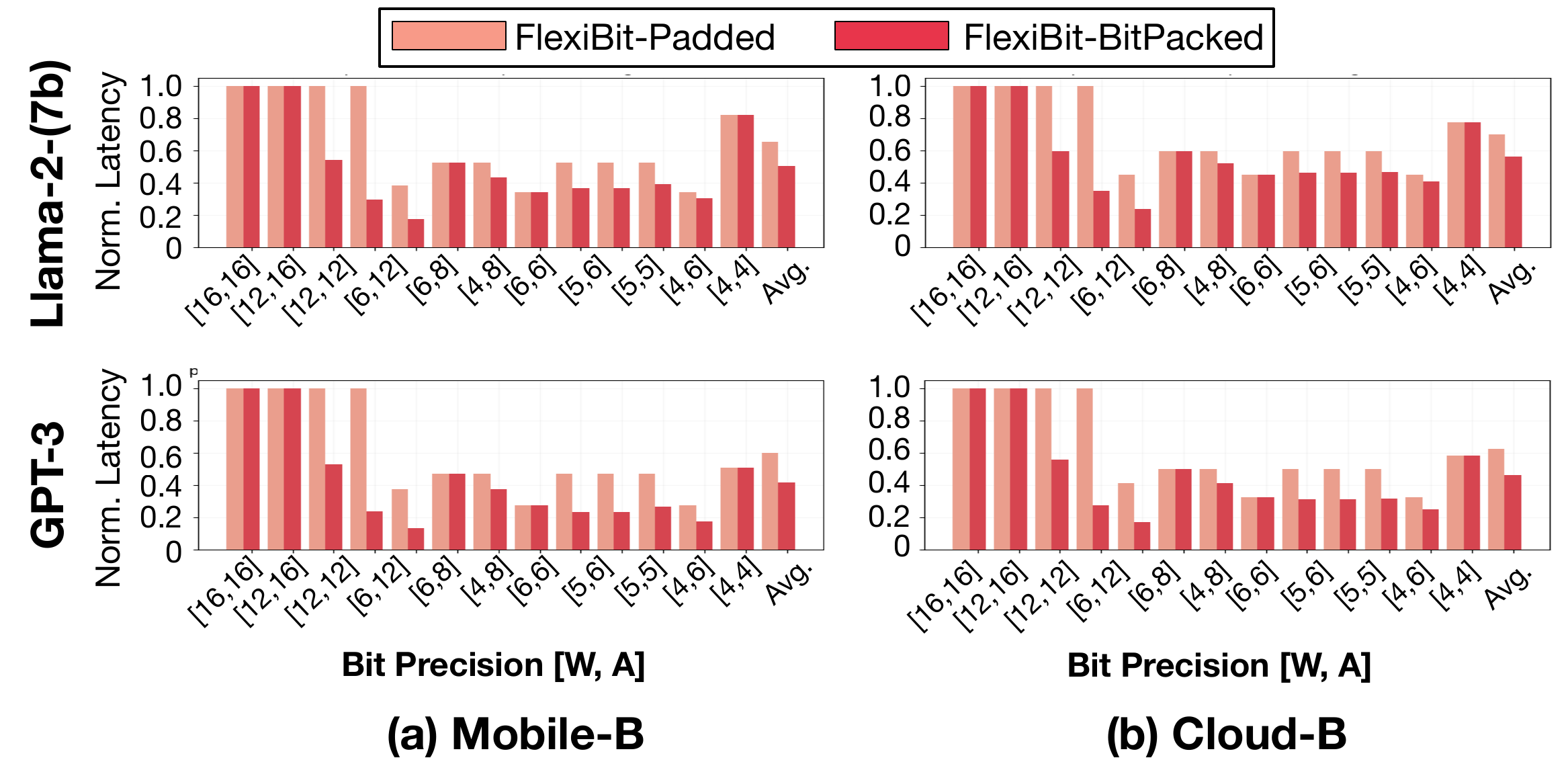}{Latency comparison of \accelerator with and without Bit-Packing structure. The results are normalized based on TensorCore Latency at each data precision.}{1}{-7}{-5}

\insertWideFigureShrink{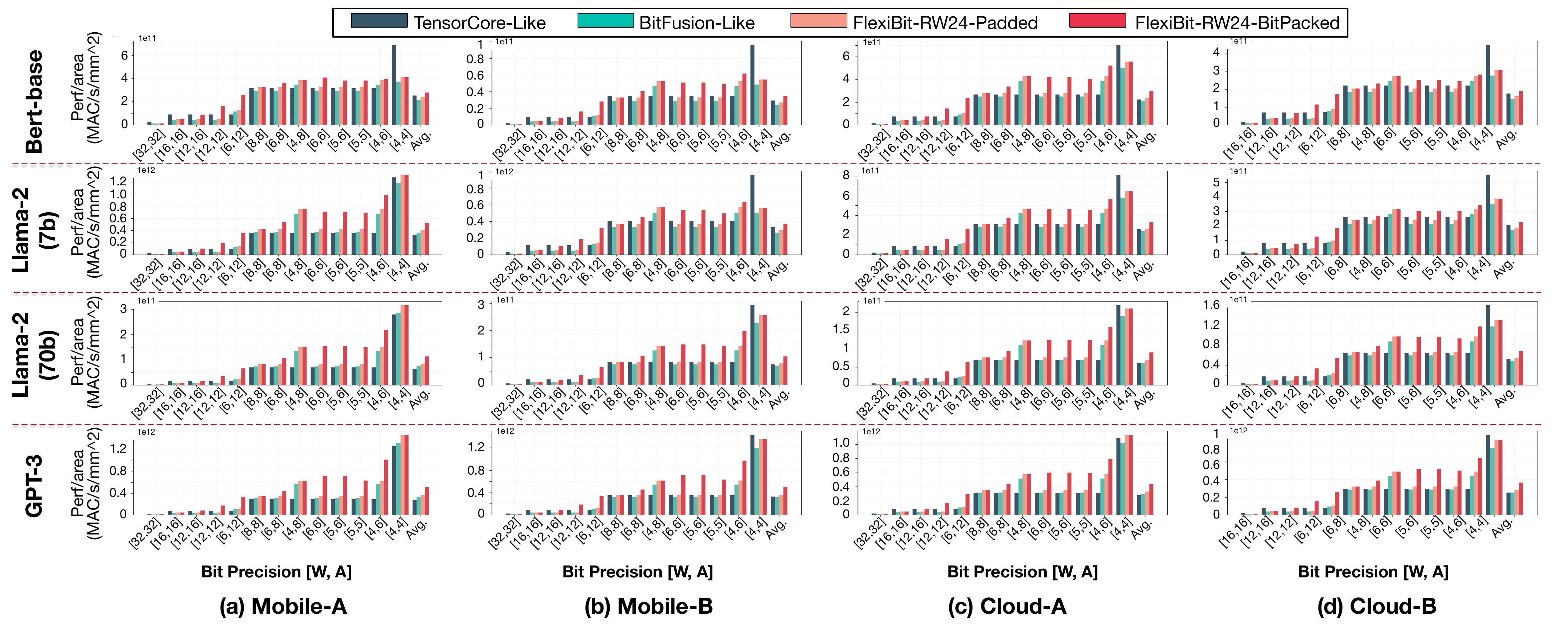}{Performance per area comparison on 13 Act./Wgt. precision pairs. The sequence length is 2048.}{1}

\insertWideFigureShrink{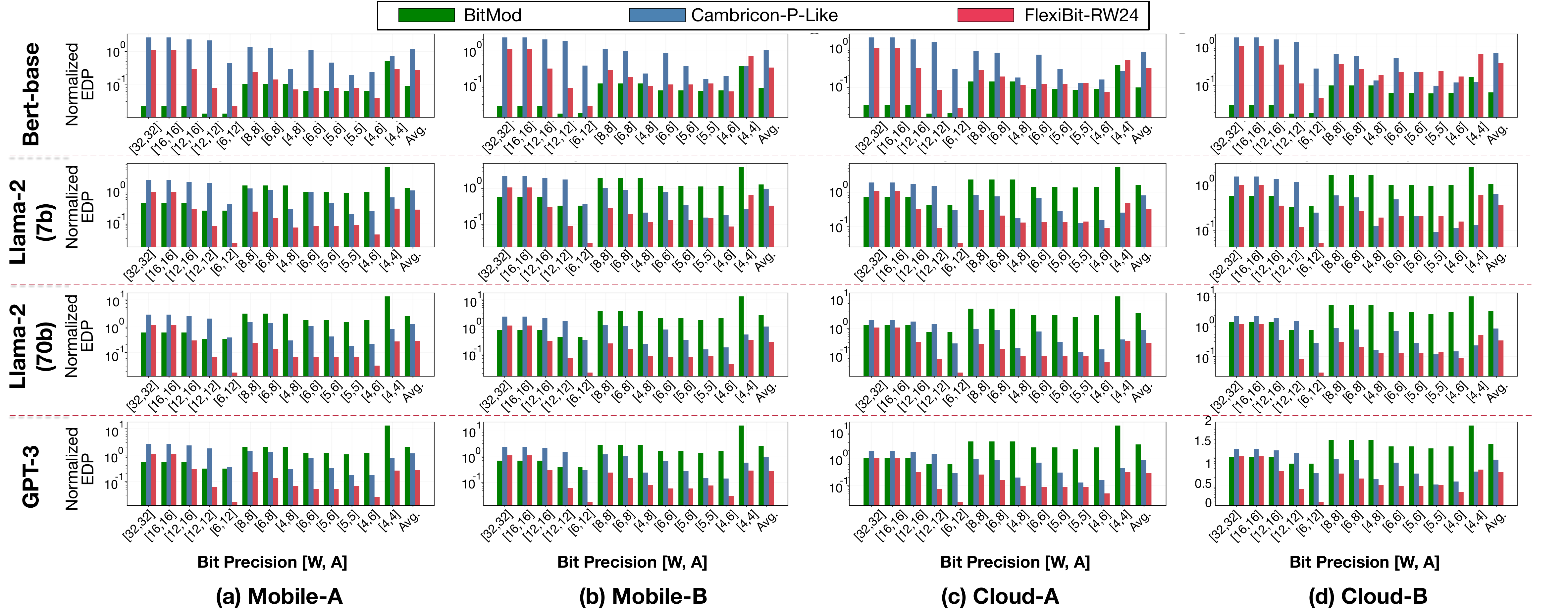}{Comparison of a bit-serial (Cambricon-P and BitMod) and our bit-parallel (\accelerator) flexible precision architecture using EDP normalized against Tensor Core-like baseline.}{1}

\insertFigureScaleTight{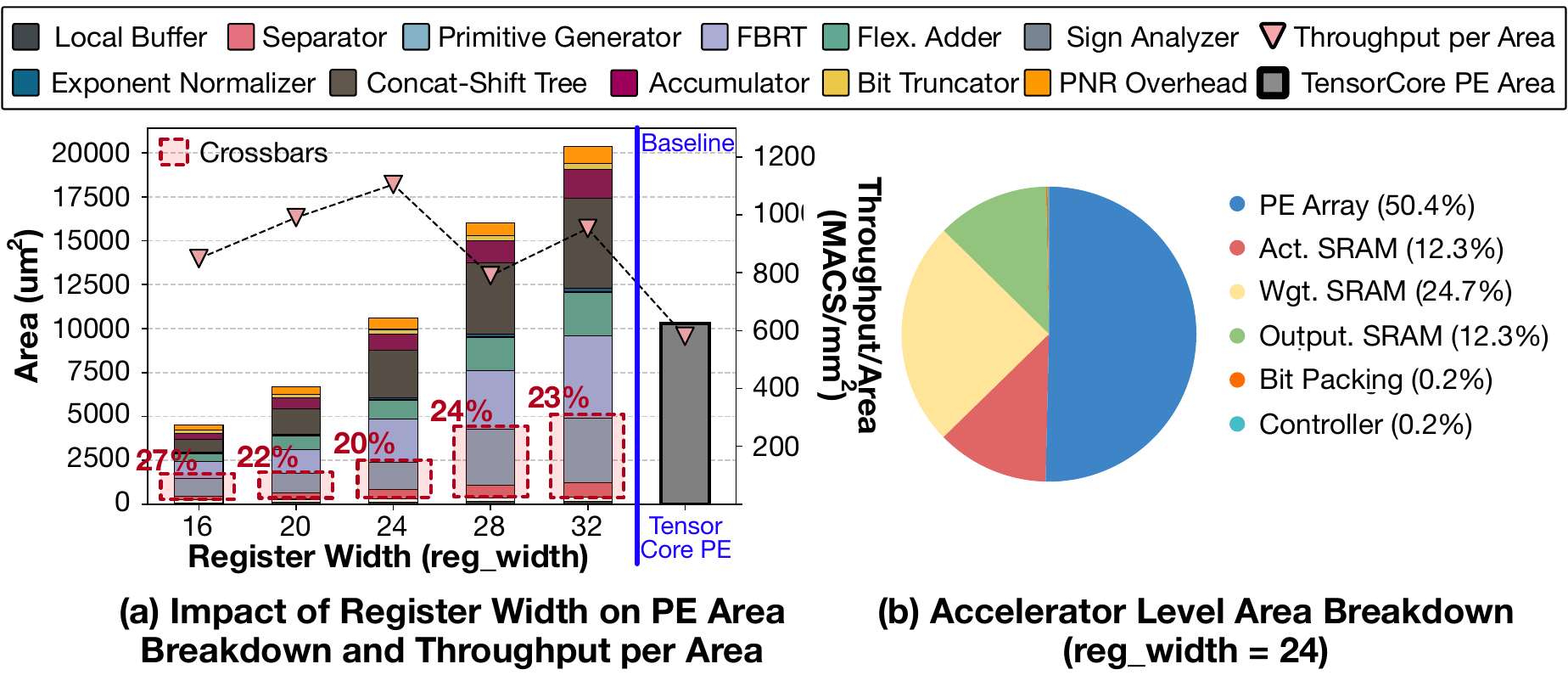}{The impact of register width and area breakdown. (a) PE Area breakdown of \accelerator and throughput per area on the register width from 16 to 32 on Mobile-A scale (b) accelerator area breakdown (Reg\_Width = 24)}{0.9}

\subsubsection{Performance Per Area}
\label{subsubsection:perf_per_area}

To understand the relative benefits over costs, we evaluate the performance per area metrics and present the results in~\autoref{fig:Perf_Area}. As we can observe, \accelerator provides better performance per area, 28\% and 34\% higher than Tensor Core and Bit-Fusion on average, respectively. The results indicate that benefits from flexible precision enabled by \accelerator exceed hardware costs for supporting them. However, TensorCore slightly outperforms~\accelerator in some power-of-two cases (i.e. [8, 8] and [4, 4]). We note that the main strength of~\accelerator is non-power-of-two precision computation, while it provides similar throughput for power-of-two precisions.



\subsubsection{Comparison against Bit-serial Accelerator}
\label{subsubsection:bit_serial_comp}

We select Cambricon-P\cite{cambricon-p} and BitMod \cite{bitmod} as the state-of-the-art bit-serial architectures, which provide flexibility in bit precisions to some extent. We also utilize BitMod~\cite{bitmod} simulator fo simulating BitMod with various precisions on weights. To consider both costs and benefits, we compare \accelerator, Cambricon-P, and BitMod using Energy-Delay Product (EDP) in \autoref{fig:Bit_Serial_EDP}. Our evaluation shows that Cambricon-P is best in power and energy-efficiency, working with 7.1$\times$ less power, on average. However, bit-serial architectures cannot achieve bit-parallel performance. Cambricon-P requires 52$\times$ more latency on Llama2-70b and Cloud-B scale due to the serialized computation for different precisions, which is a significant drawback. This led to 2.48$\times$ more EDP over Cambricon-P compared to \accelerator on Llama-2-70b, demonstrating that the benefits of \accelerator exceed the costs. Compared to BitMod, our accelerator provided 7.9$\times$ better latency. The significant latency differences originate from BitMod's fixed precision for activations, long latencies for multiplications with larger bit widths, and the limited degree of bit parallelism.~\autoref{tab:EDP} and ~\autoref{tab:Area_Power} includes the performance and architecture details of all three accelerators.
However, BitMod provides 2.7$\times$ higher energy efficiency due to its light-weight bit-serial mechanism. Latency and energy together, \accelerator provides 2.9$\times$ lower EDP compared to BitMod to run Llama-2-70b model, which demonstrates that \accelerator is a promising solution. In particular, the high performance is crucial for recent large-scale models, which makes \accelerator a future-proof architecture.


\begin{table}[t]
\centering
\small
\caption{Average latency, energy, and EDP of different scales and accelerators on Llama-2-7b and Llama-2-70b models.}
\vspace{-2mm}
\resizebox{\columnwidth}{!}{
\begin{tabular}{|c|c|cc|cc|cc|}
\hline
\multirow{2}{*}{\textbf{Scale}} & \multirow{2}{*}{\textbf{Accel.}} 
& \multicolumn{2}{c|}{\textbf{Latency ($s$)}} 
& \multicolumn{2}{c|}{\textbf{Energy ($\mu J$)}} 
& \multicolumn{2}{c|}{\textbf{EDP ($\mu J.s$)}} \\
\cline{3-8}
 & & \textbf{Llama2} & \textbf{Llama2} 
   & \textbf{Llama2} & \textbf{Llama2} 
   & \textbf{Llama2} & \textbf{Llama2} \\
 & & \textbf{7b} & \textbf{70b} 
   & \textbf{7b} & \textbf{70b} 
   & \textbf{7b} & \textbf{70b} \\
\hline
\multirow{3}{*}{Mobile-B}
    & Cambricon-P\cite{cambricon-p} & 83.95 & 1195.55 & 0.54 & 7.62 & 45.33 & 9110.09 \\
    & BitMod\cite{bitmod}           & 6.94 & 151.12 & 2.99 & 36.18 & 20.75 & 5467.52 \\
    & FlexiBit                      & 1.52 & 20.52 & 9.84 & 135.86 & 14.95 & 2787.84 \\
\hline
\multirow{3}{*}{Cloud-B}
    & Cambricon-P\cite{cambricon-p} & 20.58 & 249.28 & 0.38 & 4.65 & 7.82 & 1159.15 \\
    & BitMod\cite{bitmod}           & 1.73 & 37.78 & 2.99 & 36.18 & 5.17 & 1366.88 \\
    & FlexiBit                      & 0.45 & 4.78 & 8.92 & 97.70 & 4.01 & 467.01 \\
\hline
\end{tabular}
}
\vspace{-3mm}
\label{tab:EDP}
\end{table}

\begin{table}[t]
\centering
\small
\caption{Area and Power comparison with Mobile-A scale.}
\vspace{-3mm}
\begin{tabular}{|c|c|c|c|}
\hline
\textbf{Scale} & \textbf{Accel.} & \textbf{Area ($mm^2$)} & \textbf{Power (mW)} \\
\hline

Mobile-A & Cambricon-P\cite{cambricon-p} & 5.11 & 122.15 \\
Mobile-A & BitMod\cite{bitmod} & 4.70 & 629.76 \\
Mobile-A & FlexiBit            & 18.62 & 873.48 \\
\hline
\end{tabular}
\label{tab:Area_Power}
\end{table}

\subsubsection{Area Analysis}
\label{subsubsection:area_analysis}

We present the area breakdown of \accelerator and its PE in~\autoref{fig:Area_Breakdown} based on the Mobile-A configuration. In the accelerator-level, we observe the negligible overhead of Bit-packing unit. Bit-packing implementation includes a base with 64-to-64 crossbar accompanied by an indexing mechanism for a 64-bit off-chip channel. For larger channel widths, area overhead doesn't exponentially increase since we duplicate the base implementation (e.g., two base bit-packing units for 128-bit channel). We measure the same routing overhead as TensorCore in accelerator-level(12\%), since~\accelerator is very similar to typical ML accelerators in high-level. In the PE-level, however, there is 6\% routing and wiring overhead. We observe that core modules for flexible precision, \ourtree and Primitive Generator, account for about 50\% of PE area.

\noindent
\betterparagraph{Determining the design parameter} We conduct an area analysis of \accelerator's PE varying the key design parameter, $reg\_width$. As the results in~\autoref{fig:Area_Breakdown}(a) show, larger $reg\_width$ super-linearly increase. However, the relative overhead of crossbars does not meaningfully change. To determine proper $reg\_width$, we consider both area and performance utilizing throughput per area. As shown in~\autoref{fig:Area_Breakdown}(a), we observe the best result for $reg_width = 24$, which drove our design choice.

\section{Related Works}
\label{sec:related_works}

\begin{table}[t]
\centering
\caption{Architecture categories and their readiness for LLM requirements with flexible bit FP/INT quantization.}
\vspace{-3mm}
\scriptsize
\resizebox{0.8\columnwidth}{!}{
\begin{tabular}{|c|c|c|c|c|c|}
\hline
\multirow{2}{*}{\textbf{Architecture}}
& \textbf{FP Precision/Format}
& \multirow{2}{*}{\textbf{High-Performance}}
& \multirow{2}{*}{\textbf{Scalability}}
\\
& \textbf{Flexibility}
& 
&
\\
\hline
Bit-serial  & \multirow{2}{*}{\checkmark}  &  &  \\
\cite{judd2016stripes,eckert2018neural,delmas2019bit}  &   &  &  \\
\hline
Fixed Precision/Format  &   & \multirow{2}{*}{\checkmark} & \multirow{2}{*}{\checkmark} \\
Bit-parallel~\cite{nvidia_h100,jouppi2023tpu}  &   &  &  \\
\hline
Power-of-two  & \multirow{2}{*}{Limited} & \multirow{2}{*}{\checkmark}  & \multirow{2}{*}{\checkmark}  \\
Bit-parallel~\cite{sharma2018bit}   &   &  &  \\
\hline
Precision/Format Preset  & \multirow{2}{*}{Limited} & \multirow{2}{*}{\checkmark} &   \\
flexible Bit-parallel~\cite{rapid}  &  & &  \\
\hline
Fully flexible  & \multirow{2}{*}{\checkmark} & \multirow{2}{*}{\checkmark} &  \multirow{2}{*}{\checkmark} \\
Bit-parallel (This Work)  &   & &  \\
\hline
\end{tabular}
}
\label{tab:related_work_table}
\vspace{-5mm}
\end{table}

\betterparagraph{Non-trivial Precision in Recent AI Models}
Recent AI research has witnessed a notable rise in the adoption of non-trivial precision and format. For example, \cite{micikevicius2022fp8} proposed more compact FP8 models while demonstrating model performance comparable with FP16. FP6-LLM \cite{xia2024fp6} designed a GPU kernel with unified Tensor Core support on FP6 format that shows $1.69\times$-$2.65\times$ higher normalized throughput on LLaMA-70b than FP16. ANT \cite{guo2022ant} proposed a fixed-length adaptive numerical data type to achieve low-bit quantization.

\betterparagraph{Flexible Precision Accelerators}
Cambricon-P~\cite{cambricon-p} processed different bit-serial bitflows in parallel,  with bit-indexed Inner-Product Units to address the large dependency chains that cause the low hardware utilization.
BitMoD~\cite{bitmod} introduces new precision representations that need dequantization on the fly. It uses multiple bit-serial multiplication lanes to perform W4A16 matrix multiplication and dequantization. 
However, bit-serial architectures often suffer from limited performance when targeting a large-scale system.
Also, Cambricon-P's main target is scientific applications with a relatively small number of operands (compared to deep learning) with very long precision ($10^{3}$ - $10^{5}$ bits), while deep learning typically uses billions of less than 16-bit data.
Bit-fusion \cite{sharma2018bit} proposes fusible units for integer multiplication operation that enables utilizing all the logic for combinations of power-of-two operands (e.g. A8W4).
Mix-GEMM \cite{mix_gemm} is another integer-only flexible-precision accelerator.
RaPiD \cite{rapid} supports five different FP/INT precision/formats (FP16, FP8-e4m3, FP8-e5m2, FP9, INT4, and INT2).
Although RaPiD introduces a novel architecture that maximizes usage of hardware components, the flexibility is limited to the five different presets.
Following RaPiD, All-rounder \cite{noh2023all} supports more data precisions/formats. However, its performance degrades due to the costs for more flexibility.
\cite{2_in_1_Accelerator} and \cite{Lebedev2014SpeedingupCN} support dynamic precision level adjustment with a trade-off between latency and accuracy.
However, those works do not fully exploit precision adjustments for all types of AI models and tasks.
As summarized in~\autoref{tab:related_work_table}, only this work satisfies all the desired features for large language models with flexible FP/INT precision/format data.


\section{Conclusion}
\label{sec:conclusion}

Algorithmic research in AI domain continuously reports the potential of non-power-of-two quantization. However, their benefits were sealed by the limitation of current hardware that only supports power-of-two precisions in standard formats.
Therefore, in this work, we explored a new accelerator architecture to eliminate the constraint from hardware that hinders exploration of non-power-of-two precisions in AI models. Our work shows that the benefits of flexible architecture can exceed the costs, which motivates further exploration of this direction.



\bibliographystyle{ACM-Reference-Format}
\bibliography{ref}

\end{document}